\documentclass[sn-nature]{sn-jnl}

\usepackage{graphicx}%
\usepackage{multirow}%
\usepackage{amsmath,amssymb,amsfonts}%
\usepackage{amsthm}%
\usepackage{mathrsfs}%
\usepackage[title]{appendix}%
\usepackage{xcolor}%
\usepackage{textcomp}%
\usepackage{manyfoot}%
\usepackage{booktabs}%
\usepackage{algorithm}%
\usepackage{algorithmicx}%
\usepackage{algpseudocode}%
\usepackage{listings}%

\newcommand{\apjl}{Astrophys. J. Lett.}   
\newcommand{\apjs}{Astrophys. J. Suppl. Ser.}   
\newcommand{\ao}{Appl. Opt.}   
\newcommand{\aap}{Astron. Astrophys.}   

\begin{document}

\title
{
Pulsar radio emission from thunderstorms and raindrops of particles in the magnetosphere
}

\author[1,2]{\fnm{X.} \sur{Chen}}
\author[1,2]{\fnm{Y.} \sur{Yan}}
\author*[1,2,3]{\fnm{J.~L.} \sur{Han}}\email{hjl@bao.ac.cn}

\author[1,2,3]{\fnm{C.} \sur{Wang}}

\author[1,2,3]{\fnm{P.~F.} \sur{Wang}}

\author[1,2]{\fnm{W.~C.} \sur{Jing}}
\author[4,5]{\fnm{K.~J.} \sur{Lee}}
\author[6,7]{\fnm{B.} \sur{Zhang}}

\author[4,5]{\fnm{R.~X.} \sur{Xu}}
\author[1,2]{\fnm{T.} \sur{Wang}}
\author[1,2]{\fnm{Z.~L.} \sur{Yang}}

\author[1,2]{\fnm{W.~Q.} \sur{Su}}
\author[1,2]{\fnm{N.~N.} \sur{Cai}}
\author[2,4,5]{\fnm{W.~Y.} \sur{Wang}}

\author[2,4]{\fnm{G.~J.} \sur{Qiao}}
\author[1,3]{\fnm{J.} \sur{Xu}}
\author[1,2]{\fnm{D.~J.} \sur{Zhou}}

\affil[1]{\orgdiv{National Astronomical Observatories}, \orgname{Chinese Academy of Sciences}, \orgaddress{\city{Beijing}, \postcode{100101}, \country{China}}}

\affil[2]{\orgdiv{School of Astronomy and Space Sciences}, \orgname{University of Chinese Academy of Sciences}, \orgaddress{\city{Beijing}, \postcode{100049}, \country{China}}}

\affil[3]{\orgdiv{CAS Key laboratory of FAST, National Astronomical Observatories}, \orgname{Chinese Academy of Sciences}, \orgaddress{\city{Beijing}, \postcode{100101}, \country{China}}}

\affil[4]{\orgdiv{Department of Astronomy}, \orgname{Peking University}, \orgaddress{\city{Beijing}, \postcode{100871}, \country{China}}}

\affil[5]{\orgdiv{Kavli Institute for Astronomy and Astrophysics}, \orgname{Peking University}, \orgaddress{\city{Beijing}, \postcode{100871}, \country{China}}}

\affil[6]{\orgdiv{Nevada Center for Astrophysics}, \orgname{University of Nevada}, \orgaddress{\city{Las Vegas}, \postcode{89154}, \country{USA}}}

\affil[7]{\orgdiv{Department of Physics and Astronomy}, \orgname{University of Nevada}, \orgaddress{\city{Las Vegas}, \postcode{89154}, \country{USA}}}

\abstract{
Pulsars radiate radio signals when they rotate.  However, some old pulsars often stop radiating for some periods. The underlying mechanism remains unknown, while the magnetosphere during nulling phases is hard to probe due to the absence of emission measurement. Here we report the detection and accurate polarization measurements of sporadic weak narrow dwarf pulses detected in the ordinary nulling state of pulsar B2111+46 via the Five-Hundred-Meter Aperture Spherical radio Telescope (FAST). Further analysis shows that their polarization angles follow the average polarization angle curve of normal pulses, suggesting no change of magnetic field structure in the emission region in the two emission states. Whereas radio emission of normal individual pulses is radiated by a thunderstorm of particles produced by copious discharges in regularly formed gaps, dwarf pulses are produced by one or a few raindrops of particles generated by pair production in a fragile gap of this near-death pulsar.
}

\maketitle

How and why pulsars radiate has remained elusive since their discovery over 50 years ago. In general, a pulsar radiates pulses continuously in every rotation period. The averaged pulse profiles often occupy a small fraction of the rotation longitude, which define the emission window\cite{Manchester+1995}. Analysis of the pulse polarization properties suggested that radio emission is generated by highly relativistic particles streaming in the open magnetic field lines footed on the polar cap\cite{Radhakrishnan69,rs75}, and the polarization angles reflect the magnetic field geometry of emission region sweeping across the line of sight\cite{okj20}. Though the averaged pulse profile of a pulsar is generally stable, individual pulses in each period show diverse variations. Some relatively old pulsars often cease radiation some periods, which is called ``nulling'' \cite{Backer1970,Ritchings1976}. 

The magnetosphere of an active pulsar is believed to be filled with a continuously replenished electron-positron plasma\cite{Sturrock1971}. 
Recent particle-in-cell simulations\cite{pts20, 2020ApJ...889...69C, cgc+21, bba22} show that gaps, electric discharge and pair production can occur in several preferable regions in the pulsar magnetosphere.
Radio emission of a pulsar can quench due to two possibilities. The first is the standard picture of pair cascade depletion due to the inadequate electric potential in the gap. The other is that a gap is flooded by a pair plasma produced and injected from elsewhere in the magnetosphere.
The magnetosphere should be in very different physical state when the emission ceases. A clear hint comes from the much smaller spin-down rates of a few  pulsars\cite{klo+06,llm+12,crc+12} during their long-term nulling state than those for the emission-on state, indicating the interplay between the pulsar braking  and {outer flowing particles}  in the magnetosphere. However, it is almost impossible to probe the the magnetosphere state when emission completely ceases. 

We detect a number of sporadic dwarf pulses (see Fig.1), i.e. the narrow weak pulses, in the mostly asymptotic emission-quenched state of B2111+46 by using the FAST. Detailed analyses of these dwarf pulses, such as the energy distribution (see Fig.2), emerging phase in rotation longitudes and the polarization properties, shedding new light on the long-standing enigma of pulsar nulling and mode switching, offering understanding of the physical proccesses in pulsar magnetospheres. 

PSR B2111+46 is a strong pulsar with a period\cite{hlk+04} of 1.0146848~s and a dispersion measure of 141.40~rad~m$^{-2}$ discovered by the Jodrell Bank telescope\cite{Davies1970}. Its radio emission shows two known states. In the normal emission state, the mean pulse profile shows three dominant components\cite{Mitra2004}: a central core component coming from the emission beam center and two shoulders from the conal emission. Two additional hidden components were revealed by model fitting\cite{Zhang2007, Thomas2010}. These prominent strong components in total occupy a longitude range of about 1/4 of the rotation period according to the mean profile of previous observations. Analyses of polarization profiles suggest that the magnetic axis is close to the rotation axis, and the line of sight cuts the emission beam along a large arc. The pulsed emission is generated from a region at several hundreds to more than a thousand kilometers above the neutron star surface\cite{Mitra2004, Zhang2007, Thomas2010}. Very impressive is the nulling state of PSR B2111+46 (see Extended Data Fig. 1 -- 4), which occurs for about 10-20\% of the total periods\cite{Ritchings1976, Gajjar2012} depending on observational frequencies. In the periods of nulling the pulsar suddenly becomes undetectable.

PSR B2111+46 is serendipitously observed by FAST in three sessions in 2020 August and September (see Table 1) during the FAST Galactic Plane Pulsar Snapshot survey\cite{Han2021}, and the verification observations for the dwarf pulses were made on 2022 March. After radio frequency interference (RFI) is removed and data calibrated, the polarization profiles for every individual period and the mean profiles for each session are obtained (see Fig.1 and also Extended Data Fig.1 -- 4). Thanks to the high sensitivity of the FAST, we detected a large number of dwarf pulses (see Fig.1 and also Extended Data Fig. 1--5) emerging occasionally from ordinary nulling periods, and their polarization properties are also well-measured. Such dwarf pulses are rare, and merely a few have previously been detected from PSR J1107$-$5907\cite{yws+14}, and non further information was previously available for further physical studies. The dwarf pulses of PSR B2111+46 are generally undetectable in the low sensitivity and/or low time-resolution observations, and hence these periods with dwarf pulses are ordinary thought to be in the nulling state. Therefore these distinctive dwarf pulses are nice probes for  physical processes and the emission region in most asymptotic quenched state of magnetosphere.

Dwarf pulses of PSR B2111+46 distinguish themselves from normal pulses by their distinctly small energies (see Fig.2). For many pulsars the energy distribution of individual pulses follows a log-normal distribution \cite{Burke2012}. The emission of PSR B2111+46 from the normal state also follows such a distribution. However, the dwarf pulses we detect are very weak and narrow. Therefore, they stay far away from the normal pulse energy distribution  (see Fig.3). Just in a sharp contrast to the giant pulses observed from some young pulsars, most dwarf pulses have lower peak spectral densities than regular pulses, while giant pulses have spectral densities typically more than one order of magnitude higher than normal pulses\cite{cstt96,Burke2012}.

Besides small energie s, the dwarf pulses detected from PSR B2111+46 also have very narrow pulse widths (see the distribution of $W$ in Fig.3). With the sampling rate of 49.152~$\mu$s each data point, for the pulsar PSR B2111+46, the FAST can measure the radio emission of more than 8870 samplings inside the emission beam (see the longitude range defined by the profile in Fig.4) of PSR B2111+46 among the totally 20643 data points every period. The normal individual pulses mostly have pulse widths in the range of $60^{\circ}<W <100^{\circ}$ with a diverse intensity fluctuations along longitudes, as if in the ``thunderstorm mode'' combined by a huge number of emission ``cells" (see Fig.1 and also two more examples in Extended Data Fig.5), while the dwarf pulses consist of only one (see the period No. 237 in Fig.1) or only a few resolved peaks (see Extended Data Fig.5) as if one or few raindrops in the clear sky, with each elementary pulse of about 0.1$^{\circ}$ (about 0.3ms). Such a time scale is much shorter than the classic subpulses but much longer than micro-pulses that have a time-scale of ns or $\mu$s\cite{rhc+1975,spb+2004}. The dwarf pulses can appear across a wide range of phases for both the core and conal components and even in between, with a preference in the trailing component (see Extended Data Fig.6).

Polarization measurements provide a physical link between the detected emission and the magnetic field lines in the emission region\cite{okj20}. PSR B2111+46 has a grand-designed ``S"-shape polarization angle (PA) curve for the mean linear polarization profile, which has been used to estimate the emission height and swept-back of magnetic field lines for the central emission components\cite{Mitra2004}. From our sensitive observations, we found an much extended leading wing and the orthogonal mode for weak conal emission wings in both leading and trailing longitudes (see Fig.4). In some periods, radio emission is detected only for one or two of the three main components (see Extended Data Fig.1 -- 4) which corresponds to partial nulling\cite{Gajjar2012}. The most intriguing fact is that polarization angles of the dwarf pulses, together with the partially nulling pulses, all nearly follow the PA curve of the mean profile or at the respective orthogonal mode (see Fig.4). The detection of dwarf pulses in the ordinary nulling state from PSR B2111+46 that still keep the same polarization properties as normal pulses suggests that the magnetic field configuration does not change at the transition phase to the completely nulling phase. 

How and where are these dwarf pulses generated in such ordinary nulling periods? Why does the radio emission of PSR B2111+46 often cease?  
The nulling state reflects the deficient of outer-flowing particles for radiation, or the failure of the coherence condition for particles, or even quenched gaps by flooding pairs formed in other parts in the pulsar magnetosphere.
PSR B2111+46 has a characteristic age of $2.25\times 10^{7}$~year and the surface magnetic field of $8.62\times10^{11}$~Gauss and is located in the death valley in the $P- \dot P$ diagram (see Extended Data Fig.7).  
The pair creation of such a pulsar can operate effectively only above the magnetic polar cap\cite{2020ApJ...889...69C,cgc+21} through the $\gamma-B$ process where the field is strong enough. 
For such an old pulsar with a weak magnetic field, the gap voltage is often barely enough to ignite electron-positron discharges, so that a pulsar may fail to radiate from time to time.

If dwarf pulses are generated by one or few raindrops of streaming particles from the otherwise nulling state, it means that only one or a few lightnings ignite above the polar cap so that a barely formed gap is very quickly discharged. Our observations shown in Fig.5 indicate that the spectra ($S \sim {\nu}^{\alpha}$) of some distinguishable emission components are various, with a possible index $\alpha$ from $-$5 to the unexpectedly +5, and the dwarf pulses is more likely to have a reversed spectrum (see Extended Data Fig.8).  
Normal individual pulses with many distinguishable peaks, revealed by the FAST observations in Fig.1 and Extended Data Fig.5, indicate that the lightnings, pair-creation cascades and related physical processes occur in a very wide area of the polar cap, as if the emission is produced by a thunderstorm of particles. The phase-resolved spectra are more likely to be flatter or even reversed in the two-side conal phase-ranges (see Extended Data Fig.9).
 
The plasma properties in the magnetosphere can be examined by the propagation effects \cite{cr79,wlh10,bp12}, such as adiabatic walking and polarization limiting radius. The density changes in the nulling state could cause the polarization angle curve shifted to an earlier or later rotation phase with an extent depending on the background plasma properties and magnetic field strength. The longitude shift of polarization angle curves of the dwarf pulses from that of normal pulses is found to be $-0.77^{\circ}\pm0.25^{\circ}$ (see Extended Data Fig.10) from our FAST measurements for PSR B2111+46, which is marginally significant and implies not only no change on magnetic field configuration in emission region, {but also only a slight change or no change ($22\pm7\%$) of the density of magnetospheric background plasma in the nulling state, compared with that for the normal emission state.} 

In addition to PSR B2111+46, dwarf pulses have also been detected from some nearly-nulling periods of several other pulsars by FAST observations, such as PSRs J0540+3207, PSR J1851$-$0053 and J1946+1805. A small number of narrow pulses previously detected from PSR B1237+25\cite{zr05} are similar to dwarf pulses presented here.  Dwarf pulses are likely a common phenomena for old nulling pulsars, a distinct very weak emission mode\cite{yws+15} {standing out more clearly in observations with a higher sensitivity}. Detailed high-time resolution polarization observations of dwarf pulses as here in this paper can promote further our understanding of the radiation mechanism of radio pulsars.

\section*{Methods}
\subsection*{FAST observations of PSR B2111+46}\label{sect-M1}

PSR B2111+46 was serendipitousely detected in one of the 19 beams of the L-band 19-beam receiver on August 24, August 26, and September 17, 2020 while the FAST was tracking other objects for verification of pulsar candidates from the FAST Galactic Plane Pulsar Snapshot survey\cite{Han2021}. Each tracking observation lasts for 15 minutes, see Table 1 for details, i.e. 885/886 periods of PSR B2111+46. On March 8, 2022, the central beam of the L-band 19-beam receiver of the FAST was focused on PSR B2111+46 for 2 hours, without beam offset, to verify the detection of dwarf pulses.

In all observations, the signals from X and Y polarization channels in the radio frequency range of 1.0 to 1.5 GHz are amplified and then transferred to the digital room via the optical fibers. Radio frequency signals are recovered and sampled, and then channelized to 2048 channels in the digital backend, and then composited to 4 polarization for each channel, XX, YY and X*Y and XY*  (see details in \cite{Han2021}). These polarization data are accumulated 49.152~$\mu$s (as the sample rate for these 4 polarization channels) and recorded into a set of fits files. For each session, we have 2 minutes observations before the session with calibration signals of an amplitude of 1~K switching on-off every 1 second. This part of data are processed to form a calibration reference file, which is used to calibrate polarization channels.   

\subsection*{Data processing}  \label{sect-M2}

The raw data of FAST observations of PSR B2111+46 are all saved in a search mode, with the 4 polarization channels all recorded every 49.152~$\mu$s. Based on the pulsar ephemeris obtained from the ATNF pulsar catalog\cite{mhth05}, we process the pulsar data by using the package, DSPSR\cite{dspsr}. The data are de-dispersed according to the DM value $DM=141.26~ {\rm pc~ cm^{-3}}$ initially\cite{hlk+04}, and then are folded according to the period $P=1.0146848$~s. A better DM value $DM=141.378 ~{\rm pc~cm^{-3}}$ is found by using our high time resolution data for the sharp peaks of individual pulses. The polarization leakages are calibrated\cite{whx+23}, and the band distortion is corrected according to the {calibration reference file} obtained from the two minutes calibration on-off data. Some frequency channels with strong radio frequency interference were weighted to zero by using the software, PSRZAP\cite{psrchive}. The polarization data of all channels are then rotation-measure corrected according to the known RM value of $RM=-218.7 ~{\rm rad~m^{-2}}$ \cite{fdr15} by using the pulsar processing program, PAM\cite{psrchive}. After the data from all frequency channels are integrated, the four stokes parameters (I/Q/U/V) were then saved for 512 bins each period for the normal detection of nulling pulses. We worked out also the profiles with 1024, 2048, 4096 or 2,0643 bins, and noticed that these with 512 bins are the best for detecting the dwarf pulses.

\subsection*{Pulse profiles and polarization} 

For each session, the mean profile of PSR B2111+46 is obtained (see Extended Data Fig.1, 2, 3  and 4) after individual pulses from all periods are averaged. No recognized difference is found between the three polarization profiles (see Fig.4 for 2020 sessions, and they are consistent with results at 610 and 1408 MHz\cite{Lyne1988,Gould1998mn} and 1500 MHz\cite{Force2015} after the opposite definition of circular polarization is considered. However, much more extended profile wings are detected in the targeted verification observations in 2022-03-08 that has a much better sensitivity due to the targeted good pointing. These results indicate the excellent performance of polarization measurements for the L-band 19-beam receiver, even when the object is well-off of the beam center. In the results, a number of periods occasionally having radio frequency interference (see Table 1 have been cleaned and marked by the dashed line in the pulse stacks of individual pulses, such as the period No. 683, 684, 694 and 695 in the session of 2020-08-24. 

Though the mean pulse profiles show triple components for both cone and core emission, with a strong linear polarization for almost all longitudes except for these in the two edges. The Gaussian fittings to the mean profile always give 5 components\cite{Zhang2007}. The new observations of 2022-03-08 show two highly polarized prefix mean profile components (see Fig.4), so that the mean profile has a wide longitude range of more than $155^{\circ}$. The reversed sense of circular polarization at the center of mean profiles is the indication of the core nature of the central component\cite{Lyne1988, ran93}. The polarization angle curves follow a grand-designed ``S" shape\cite{Force2015, Mitra2004} which can be well interpreted by the rotating vector model\cite{Radhakrishnan69}. The orthogonal modes of the PA distributions are revealed by our FAST observations from the conal and two newly detected prefix wing emission components shown in Fig.4. The smoothly changed PA curve  extends in the two sides of mean profiles and smoothly varies for more than 220$^{\circ}$. Our fitting to the PA curve suggests that PSR B2111+46 is an aligned rotator, i.e. with a small inclination angle of only 6.3$^{\circ}$ between the magnetic axis from the rotation axis, and the line of sight impacts radio emission beam only 0.7$^{\circ}$ below the magnetic axis. The line of sight impacts the emission beam in about 40\% of a period, and FAST gets more than 8750 independent samplings of the emission beam among the 20643 data points every period. 

The polarization profiles of individual pulses in high time resolution (see Extended data Fig.5) can reveal many details about emission. Thanks to the extreme sensitivity of FAST observations, individual pulses frequently contain numerous peaks, indicating real variations in emission. These fine subpulses are considered to be elementary emission cells and have a much smaller width than conventional subpulses. An example of a highly isolated elementary pulse can be seen in No. 237 of Fig.1, which is a dwarf pulse and exhibits nearly 100\% polarization. 
The polarization angles of such dwarf pulses mostly follow the PA curve of mean profile, as seen in Fig.4.
While the position angle (PA) values of most elementary pulses {of normal individual pulses} are also conform to the PA curve of the mean profile, with deviations occasionally observed at various longitudes likely due to the overlaps of orthogonal modes. More intriguing is the sense change of circular polarization for some elementary pulses not near the centre of core component but in some other longitudes even of conal components (e.g. pulse No. 700 and 679 in Extended Data Fig.1, pulse No. 354 in Extended Data Fig.2, pulse No. 137 in Extended Data Fig.3, and also pulse No. 1551 in Extended Data Fig.5). This challenges the simple geometrical explanation for circular polarization\cite{rr90,hmxq98,ghw21}.   

\subsection*{Nulling and partial nulling} 

Nulling phenomenon is often observed for pulsars near the death-line in the P-Pdot diagram\cite{1993ApJ...402..264C,zhm00}. PSR B2111+46 is located in the death valley, see Extended Data Fig.7.
Nulling of PSR B2111+46 has previously been observed, and the nulling fraction is 12.5\% at 408 MHz\cite{Ritchings1976} and increases to 21\% at 610 MHz\cite{Gajjar2012}. The statistics from Table 1 for our FAST observations give a nulling factor of about 20\% at 1250 MHz. 

The nulling fraction varies from component to  component\cite{Gajjar2012}. Partial nulling of PSR B2111+46 was previously suggested\cite{Gajjar2012}, and our high-quality FAST data clearly manifest the phenomenology (see examples in Extended Data Fig.1 -- 4). The partial nulling phenomenon means that only one or two mean profile emission components exist without the rest components.  Based on our very sensitive FAST observations, we discover that many individual pulses have normal emission only for one or two components, clearly lack of emission of the rest mean profile components, for example pulse No. 679 in Extended Data Fig.1, pulses No. 365 and 371 in Extended Data Fig.2 and pulses No. 137 and 140 in Extended Data Fig.3. These partially nulling pulses, if they appear, have a peak flux density comparable to the normal pulses. The PA values of each bin follow the mean PA curve well. No question that partial nulling pulses have a smaller pulse width than normal pulses, typically $10^{\circ}<W < 60^{\circ}$.

\subsection*{Dwarf pulses and pulse energy distribution} 

The most fascinating features observed are the ``dwarf'' pulses, which are weak and narrow in nature (see Extended Data Fig.1 -- 4). These dwarf pulses appear across a wide range of phases for both the core and conal components and even in between, with a preference in the trailing component (as seen in Extended Data Fig.6).
To describe the narrow width of these weak pulses, the pulse width in this paper is measured at a level of 3$\sigma_{\rm bin}$ that is even slightly different each period due to different RFI cleaning, therefore much larger than transitional pulse width measured at a level of 50\% or 10\% of peak which is suitable for a single Gaussian components rather than the complicated combinations of so many strong and weak pulses for PSR B2111+46. The start and end phases of on-pulse region is defined as the leftmost and rightmost sides with three successive data points higher than 3$\sigma_{\rm bin}$. By counting the consecutive points over 3$\sigma_{\rm bin}$, we get the width of a pulse. It is possible that some narrow pulses have only one or two bins, which will be selected as real detection of a pulse only if the peak flux density is larger than  $8\sigma_{bin}$. Most dwarf pulses can be resolved in high-sampling FAST observations, as shown in Extended Data Fig.5 and therefore they are composited by a few elementary pulses, probably generated by several `raindrops’ of particles streaming in pulsar magnetosphere, instead of the `thunderstorm' of particles for normal individual pulses over a wide longitude.  

Just in contrast to giant pulses detected from some pulsars\cite{wwsr06,Burke2012} which are strong pulses with a few tens or even hundreds times of peak flux density of normal pulses, the dwarf pulses have a peak flux density in general much less than normal pulses. We checked and found that PSR B2111+46 has no giant pulses, i.e. the narrow pulse with a peak flux density a few times higher than the average. According to Extended Data Fig.6, most dwarf pulses have a peak flux density less than 50~mJy, more than 5 times weaker than the averaged peak, except a few very narrow pulses (e.g. the pulse No.237 in Fig.5) which have a high peak. We tried to define the dwarf pulse as a peak flux density less than e.g. 20\% of the peak in the averaged profile, but noticed that the peak is really bin-number dependent.

When the fluence of an individual pulse is counted by the area underneath each pulse profile, the normal pulses have an energy following the log-normal distribution, similar to other pulsars\cite{Burke2012}. The partial nulling pulses have less energy, mainly because of their lack of {some} emission  components. The dwarf pulses have the smallest energy, as shown in Fig.2, 
so that they are hidden in the energy distribution peaks for nulling, {but}  distinctly stand away from the log-normal distribution of normal pulses. { If observations were made with a better sensitivity, i.e. with a much smaller $\sigma_E$ in Fig.2, these dwarf pulses would stand out clearly from the histogram peak for nulling periods. }
Though some dwarf pulses have  previously been detected from PSR J1107$-$5907\cite{yws+14}, the FAST data here for PSR B2111+46 show the dwarf pulses as a distinct population for the first time. Their distinctive distribution in the two dimensional plot of pulse-width and pulse energy in Fig.3 suggest that they belong to a new class of pulses for weak emission mode\cite{yws+15}.

Combining the energy and width information, dwarf pulses have a pulse width narrower than $15^{\circ}$ and the fluence $E<2~{\rm Jy ~ms}$ (see Fig.3), reside at the lowest ends of the distribution in a separate island from the main pulses. This differs from the general mode-changing pulses\cite{Burke2012} that have a similar pulse-width distribution as main pulses or the sparse pulses in the RRAT mode of PSR J0941$-$39\cite{bb10} and PSR B0826$-$34\cite{eamn12} which have similar peak flux densities of the normal pulses.

\subsection*{Possible physical processes for different emission modes}

PSR B2111+46 exhibits four emission modes: the normal pulse mode consisting of many small, distinct elementary emission cells and some wide, undistinguished emission components; the partial nulling mode that lacks emission in some components; the dwarf pulse mode characterized by one or only a few emission cells; and the completely nulling mode. The four modes of radio emission should respond to different physical state in the magnetosphere. 

In the conventional picture, a ``gap'' with charge density below the Goldreich-Julian density\cite{gj69} is believed to be produced near the polar cap region, either in the form of a vacuum gap\cite{rs75} or space-charge-limited floe\cite{as79}, or formed in the outer magnetosphere beyond the null-charge surface\cite{chr86} or in the annular region\cite{qlw+04} extending from the surface to outer magnetosphere in the form of a slot gap\cite{mh04}. Pulsar radio emission is coherently produced by a bunch of particles, as indicated by its extremely high brightness temperature, the coherency must be realized by ordering particles in phase by the longitudinal electrostatic waves or by ``antenna mechanism”. A clump of relativistic particles streaming along a bunch of magnetic field lines can produce visible radio emission at a given frequency band from a finite height regions. The lower frequency emission is generally generated from  a higher altitude region in the pulsar magnetosphere. 

The most probable region of gap formation for this old pulsar with such a weak magnetic field, however, should be above the polar cap, as shown by recent simulations\cite{cgc+21,2020ApJ...889...69C}, which converges to the conventional concepts for the inner gap and the cascades of pairs via $\gamma-B$ precess\cite{Sturrock1971,rs75}. The electron-positron discharge near the pulsar polar cap is non-stationary \cite{pts20,cgc+21}, which leads to large amplitude fluctuations of the electric field and collective plasma motions. Any break of the non-stationary nature will lead to incoherency for the radiation, and then the emission would be very weak even through the particles are still flowing out along the field lines.

The nulling state of pulsar emission demonstrates either the deficient of outer-flowing particles for radiation, or the coherence lost for particles, or the failure of gap formation. Another possibility is that the polar gap is flooded with pair plasma created from other gaps, so that the gap and radio emission is screened \cite{bba22}. Current observations seem to support the former possibility, especially when the pulsar is old and near the radio deathline\cite{1993ApJ...402..264C,zhm00}. 

For the emission state of PSR B2111+46, the accurate measurements of the polarization properties and the fine fluctuations for the well resolved normal pulses in Extended Data Fig.5 and the dwarf pulses in Fig.1 and Fig.5 show clear that normal pulsar emission is composited by the radiation from the thunderstorm of particles over a widely distributed area above the polar cap, with a very large multiplicity of cascades and also a higher plasma density. The dwarf pulses of PSR B2111+46 are produced by one or few raindrops of particles produced by the pair discharges, with a much lower multiplicity.

In principle the gap voltage for normal pulse emission state is higher than that for dwarf pulses, so that the pair-production multiplicity is large and that the energy distributions of the created particles may also be different. To find hints for these probable changes, we then examine the spectra of phase-resolved emission of individual pulses. As shown in Fig.5, the spectra ($S \sim {\nu}^{\alpha}$) of some distinguishable emission components are various, with a possible index $\alpha$ from $-$5 to the unexpectedly +5. In the phase range for the positive spectral index, the PA values are firmly following the average PA curve. For a given dwarf pulse, the spectra do not vary along the pulse phase in such a narrow phase range. The distributions of mean spectral indexes for three kinds of individual pulses, normal pulse, partial nulling pulses and dwarf pulses, are shown in Extended Data Fig.8. Dwarf pulses most likely have a reversed spectrum with a positive index, which means the primary particles in the gap may be responsible for Dwarf pulses.   
Recent numerical simulations\cite{2020ApJ...889...69C,cgc+21} show  that the core and conal components of pulsar radio emission may be preferably produced because of the proper angle between the pair production fronts and the background magnetic fields. This in principle should induce a higher probability for elementary emission in the beam center and beam edge, which is consistent with the mean profile of PSR B2111+46. The emission of much extended phase range should be caused by the curvature of magnetic fields in the edge of emission beam.  We examine the phase-resolved spectra for all normal pulses, and found that in general the spectra between the phase range of $\pm(20^{\circ}-25^{\circ})$ are flatter than those at other phases and even more likely reversed (see Extended Data Fig.9).

\subsection*{Plasma multiplicity and propagation effects in the pulsar magnetosphere}

When pulsar emission is ceased, either due to the failure of gap formation or the coherence lost of emission particles, the pulsar magnetosphere should always have pair plasma filled but with a different multiplicity. The propagation effects of radio emission in the pulsar magnetosphere should be so affected, {which can be probed by the changes of} the polarization properties\cite{wlh10,bp12}. For example, the polarization angle follows the direction of the local magnetic field due to adiabatic walking \cite{cr79} until to the polarization limiting radius, after which the natural wave mode evolution becomes non-adiabatic and the PA angle is frozen. In this case, the PA curve should be shifted to an earlier rotation phase (see eq.~5.88 of \cite{wlh10})
\begin{equation}
\phi_\mathrm{shift}\simeq -10.5^\circ (\eta/100)^{1/3}(\gamma/100)^{-1},
\end{equation}
here $\eta=N/N_\mathrm{GJ}$ the plasma multiplicity in the pulsar magnetosphere, $\gamma$ the Lorentz factor of the background plasma stream. In principle, the phase shift $\phi_\mathrm{shift}$ can be directly determined by the phase difference between the steepest position of PA curve and the center of emission profile determined by the whole open field line emission region. However, it is very hard to determine the central phase of the profile since the edges of emission region cannot be determined clearly, and therefore we cannot get the plasma multiplicity. Nevertheless, we can compare the PA curves of dwarf pulses to the mean PA curve of normal pulses, and obtain the difference of phase shifts as being $\Delta\phi_\mathrm{shift}=\phi_\mathrm{shift,dwarf-pulse} - \phi_\mathrm{shift,normal-pulse}$. Because radio emission at two states comes from almost the same region with the same field geometry, the change of plasma density can be limited by 
\begin{equation}
\Delta\eta/\eta \simeq -0.3 (\Delta\phi_\mathrm{shift}/1^\circ) (\eta/100)^{-1/3} (\gamma/100),
\end{equation}
for a given Lorentz factor $\gamma$. 

The accurate polarization measurements of dwarf pulses by sensitive FAST observations give a chance to probe the drop of plasma density in such emission-almost-quenched pulsar magnetosphere. The PA values of dwarf pulses almost follow the mean PA curve. The phase shift of PA for each dwarf pulse is obtained from the difference between the phase of the dwarf pulse and the phase for the same PA value in the mean PA curve.
Taking all phase-shift values for all dwarf pulses, and we fit the distribution with a Gaussion function, and obtain the mean shift as being $\Delta\phi_\mathrm{shift}\simeq-0.77^\circ\pm0.25^\circ$ (see Extended Data Fig.10), which is marginally significant and implies a small increase of background plasma density in the pulsar magnetosphere by an amount of $\Delta\eta/\eta \simeq (22\% \pm 7\%) (\eta/100)^{-1/3} (\gamma/100)$. Maybe the nulling of this pulsar is caused by the flooding of pair plasma to the inner gap\cite{bba22}, so that the pair-production and the following radio emission process cease. However, better measurements for more dwarf pulses are desired to give a better constrain of $\Delta\phi_\mathrm{shift}$ and then $\Delta\eta/\eta$.

\clearpage

\noindent{\bf Data availability}
Original FAST observational data are open sources after the 1-year protection for the high priority usage by observers, according to the FAST data policy. The processed data presented in this paper can be download from web-page http://zmtt.bao.ac.cn/GPPS/B2111/

\noindent{\bf Acknowledgements}
This work made use of data from the FAST. FAST is a Chinese national mega-science facility, built and operated by the National Astronomical Observatories, Chinese Academy of Sciences. 
The authors of work have been supported by the Natural Science Foundation of China: No. 11988101, 11833009 and National SKA Program of China 2020SKA0120100.

~

\noindent{\bf Author contributions}
XC and YY processed all related data and noticed dwarf pulses, and they contributed to this paper equally.
JLH supervised and coordinated the team work, pursued the nature of dwarf pulses, and took the responsibility for paper writing. 
PFW, CW, WCJ and DJZ contributed to different aspects of data processing,
TW, WYW, ZLY, WQS, NNC, JX, RXX, KJL, GJQ and BZ joined the discussions and contributed to some parts of paper-writing or plot-making.

~

\noindent{\bf Competing Interests} 
The authors declare that they have no competing financial interests.

~

\noindent{\bf Additional information}
Correspondence and request for materials should be addressed to J.L. Han.

\clearpage

\newpage

\section*{Tables}


\begin{table}[ht]
\centering
\caption{$|$ Details of FAST observations of PSR B2111+46. The table includes observation date, observation target name, FAST beam name for the pulsar detection, the offset of the pulsar location from the beam center, observation time, number of periods, number of dwarf pulses detected,  number of nullings recognized, and number of periods removed due to RFI.
}
\vspace{6mm}
{\footnotesize
\tabcolsep 2pt 
\begin{tabular}{ccccccccc}
\hline
Obs. Date    & Target Name & Beam Name & Offset & Tobs    & No. of & Period No. for & Period No. for & Peiod No. by
\\
            &         &   &  ($'$) & (min.) &  Periods &  Dwarf  & Nulling 
            & RFI-cleaned 
            \\ \hline
2020-08-24  & J2113+4642  & P1M01 &  2.1  &   15  &886    & 11  & 182  & 64
\\
2020-08-26  & J2113+4645  & P1M01 &   2.4  &  15  &886    & 7   & 180  &  0
\\
2020-09-17  & J2114+4655  & P1M12 & 2.4  &   15  & 885    & 8   & 177  &  0
\\
2022-03-08  & B2111+46  & P1M01 &  0.0  &  120  & 7098   & 149 & 1563 &  1
\\
\hline
\end{tabular}}
\label{tab_obs}
\end{table}

\section*{Figure captions} 

~

\noindent
Fig.1 $|$ FAST detection of a dwarf pulse in a series pulses of B2111+46. (a) A segment of pulse trains of PSR B2111+46 observed in the session of 2022-03-08 by the FAST, showing some emission and nulling periods; (b-d): polarization profiles of 3 individual pulses. In the lower sub-panels the total intensity $I$, linear polarization $L$ and circular polarization $V$ (with positive values for the left hand sense) are plotted in the original time resolution (49.152~$\mu$s) of FAST observations, and the polarization angles (PA) are plotted in the upper sub-panels. The dwarf pulse in the period No.237 has only one resolved emission cell, almost fully linearly polarized with a width of about $0.1^{\circ}$. Many notches of the other two pulse profiles are sensitive significant detection of real intensity fluctuations caused by emission cells with different strength. The error bar for PA is  $\pm1\sigma$. The intensity is scaled with the off-pulse fluctuations expressed by $\sigma_{\rm bin}$.

~

\noindent
Fig.2 $|$ Dwarf pulses detected in the nulling periods with very low energy. The energy distribution of 822, 886, 885 and 7097 individual pulses of PSR B2111+46 observed in four FAST observation sessions. The pulse energy $E$ (i.e. the fluence) for every period is the sum of energy of an individual pulse over the full pulse-on window defined by the mean profile. To express data quality of the observations, the distribution is scaled by the standard deviations $\sigma_E$ of the stochastic energy in the same size but pulse-off window, rather than the averaged energy $\langle E \rangle$ as in literature that is too high for FAST detected dwarf pulses. The emission state and the null state show two main peaks in the histogram. The orange part overlapping on the nulling state stands for dwarf pulses. The green curve is the best log-normal fitting for the normal emission energy distribution.
 
~

\noindent
Fig.3 $|$ Dwarf pulses of PSR B2111+46 as a distinct population from the partial nulling and normal pulses. Pulse width is measured at the most outer profile at the 3$\sigma$ detection level. {\it Upper panels}:  Pulse fluence integrated over each pulse, $E$, against pulse width, $W$, and the density distribution of data are presented in color and also in contours at levels of 1/$2^{-n}$ of the maximum density ($n$ = 1 -- 8). More sensitive observations on 2022-03-08 gives larger widths for normal pulses. No mention of about 20\% of nulling period, normal pulses are concentrated around the main peak, with a fluence in the range from 10 to about 50~Jy$\cdot$ms and a pulse width $60^{\circ}<W<100^{\circ}$. The dwarf pulses are concentrated on another peak, with a fluence less than 1 Jy$\cdot$ms and a pulse width less than $15^{\circ}$ (i.e. 40 ms). In between are partially nulling pulses (see Method: Individual pulses). {\it Lower panels}: The same as {\it the upper panel} but for peak flux density $S$ against the pulse width. The {\it left panels} are made for individual pulses obtained the three sessions in 2020, the {\it right panels} for pulses detected in the longer verification observation session on 2022-03-08.

~

\noindent
Fig.4 $|$ Polarization angle distribution of dwarf pulses compared to the data of normal pulses. The PA data of every bin of dwarf pulses (yellow) are plotted against those of normal pulses (green) and the mean polarization profiles in the 4 FAST observation sessions. The orthogonal modes are mostly predominant in  the wings of conal components. The magnified polarization profiles of 2022-03-08 illuminate the newly detected much extended leading weak profile wing. The error bar for PA is $\pm1\sigma$. The intensity is scaled with the peak value.

~

\noindent
Fig.5 $|$ Various phase-resolved spectral index for two individual pulses and two dwarf pulses. The water-fall plot for the individual pulse intensity on the phase-frequency plane ({\it upper subpanels}. The frequency channels containing RFIs have been removed), clearly demonstrating the variation of phase-resolved spectral index for individual pulses (the second upper subpanels). The polarization profiles of the pulse and the PA values (green) together with the mean PA curve (grey) are plotted in the bottom subpanels and the second bottom subpanels. The observation date and the period number of the individual pulse are marked in the bottom panel. The PA curves are fitted with the rotating vector model\cite{Radhakrishnan69}. The error bar for PA is  $\pm1\sigma$. The intensity is scaled with the off-pulse fluctuations expressed by $\sigma_{\rm bin}$.



~

\noindent
Extended Data Fig.1 $|$ The pulses of PSR B2111+46 observed by the FAST on the session on 2020-08-24. The most left panel is the train of individual pulses for 886 periods, with the mean profile shown in the bottom, and the intensity is normalized using the peak value. The total energy of every individual pulse is plotted in the immediately right, so that the energy fluctuations are seen very clearly which show the two predominate peaks for both nulling and emission states in the number distributions in the bottom. A segment of the pulse stack is shown in grey for high quality individual pulses, with significant fluctuations of profile amplitude, in which normal individual pulses {can be seen} in the period No. 702-700, 696 and 680, partial nulling in the period No. 679, and dwarf pulses of the period No. 699 and 682. The detailed polarization profiles for 4 pulses are presented in the right 4 panels, each with total intensity $I$, linear polarization $L$ and circular polarization $V$ in the bottom subpanel and $PA$ in the upper subpanel. The polarization profiles of the mean pulse are shown in dashed line in these subpanels for comparison. The error bar for PA is  $\pm1\sigma$. 

~

\noindent
Extended Data Fig.2 $|$ The pulses of PSR B2111+46 observed by the FAST on the session on  2020-08-26. The most left panel is the train of individual pulses for 886 periods, with the mean profile shown in the bottom, and the intensity is normalized using the peak value. The total energy of every individual pulse is plotted in the immediately right. A segment of the pulse stack is shown in grey for high quality individual pulses, with a dwarf pulse in the period No. 377 and partial nulling in the period No. 365. The detailed polarization profiles for 4 pulses are presented in the right 4 panels, each with total intensity $I$, linear polarization $L$ and circular polarization $V$ in the bottom subpanel and $PA$ in the upper subpanel. The polarization profiles of the mean pulse are shown in dashed line in these subpanels for comparison. The error bar for PA is  $\pm1\sigma$.

~

\noindent
Extended Data Fig.3 $|$ The pulses of PSR B2111+46 observed by the FAST on the session on 2020-09-17.
The most left panel is the train of individual pulses for 885 periods, with the mean profile shown in the bottom, and the intensity is normalized using the peak value. The total energy of every individual pulse is plotted in the immediately right. A segment of the pulse stack is shown in grey for high quality individual pulses, with a dwarf pulse in the period No. 136 and partial nulling in the period No. 137. The detailed polarization profiles for 4 pulses are presented in the right 4 panels, each with total intensity $I$, linear polarization $L$ and circular polarization $V$ in the bottom subpanel and $PA$ in the upper subpanel. The polarization profiles of the mean pulse are shown in dashed line in these subpanels for comparison. The error bar for PA is  $\pm1\sigma$.

~

\noindent
Extended Data Fig.4 $|$ The pulses of PSR B2111+46 observed by the FAST on the session on on 2022-03-08.
The most left panel is the train of individual pulses for 7098 periods, with the mean profile shown in the bottom, and the intensity is normalized using the peak value. The total energy of every individual pulse is plotted in the immediately right. A segment of the pulse stack is shown in grey for high quality individual pulses, with a dwarf pulse in period of No.5895 and two partial nullings in the period No. 5891 and 5897. The detailed polarization profiles for 4 pulses are presented in the right 4 panels, each with total intensity $I$, linear polarization $L$ and circular polarization $V$ in the bottom subpanel and $PA$ in the upper subpanel. The polarization profiles of the mean pulse are shown in dashed line in these subpanels for comparison. The error bar for PA is  $\pm1\sigma$.

~

\noindent
Extended Data Fig.5 $|$ Examples of polarization profiles for two dwarf pulses and two strong individual pulses in high time resolution. All of them are  observed on 2022-03-08 by FAST with time resolution of 49.152~$\mu$s. Polarization profiles for two strong individual pulses are shown for their elongated central part in the next panel. Each rip in the profiles is real, well-significant above the noise fluctuations. These unrepresented details indicate that the observed individual pulses are an incoherent collection of many elementary pulses generated {separately} in the magnetosphere. The error bar for PA is  $\pm1\sigma$. The intensity is scaled with the off-pulse fluctuations expressed by $\sigma_{\rm bin}$.

~

\noindent
Extended Data Fig.6 $|$ Longitude distribution of dwarf pulses. Longitude distribution of dwarf pulses locations are compared to the mean pulse profile indicated by the dash line. 
The bar length stands for dwarf pulse width, and  the dots mark the peak locations in the longitude.

~

\noindent
Extended Data Fig.7 $|$ Pulsar $P-\dot{P}$ diagram and the location of PSR B2111+46 in the death valley. The death lines are given for the curvature radiation in a dipole field (upper one) and an extremely curved field (lower one) in the vacuum gap model (sold lines) and the space-charged-limited flow model (dashed lines) given in \cite{zhm00}. All pulsar data are taken from the ATNF pulsar Catalogue\cite{mhth05} (version 1.70). The background gray dashed and dotted lines stand for constant surface magnetic field strengths and characteristic ages, respectively.

~

\noindent
Extended Data Fig.8 $|$ Distributions for spectral indexes of three kinds of individual pulses. All these pulses, including 5175 normal pulses, 199 partial nulling pulses and 67 dwarf pulses, are observed by FAST on 2022-03-08. The indexes are calculated for each individual pulse by using the on-pulse integrated intensity, and have an uncertainty less than 0.5. 

~

\noindent
Extended Data Fig.9 $|$ The number distribution of phase-resolved spectral indexes. Data of spectral indexes have an uncertainty less than 0.5 for all individual pulses observed by FAST on 2022-03-08, as shown in the upper subpanel, together with the mean polarization profile for understanding in the lower subpanel scaled with the peak value.

~

\noindent
Extended Data Fig.10 $|$ The phase shift distribution of polarization angles of 62 dwarf pulses. The shift values are obtained by comparison of their PA to the mean PA curve at the longitude of dwarf pulses.

\bibliography{main617}

\begin{thebibliography}{10}
\expandafter\ifx\csname url\endcsname\relax
  \def\url#1{\burl{#1}}\fi
\expandafter\ifx\csname urlprefix\endcsname\relax\def\urlprefix{URL }\fi
\providecommand{\bibinfo}[2]{#2}
\providecommand{\eprint}[2][]{\url{#2}}
\providecommand{\doi}[1]{\url{https://doi.org/#1}}
\bibcommenthead

\bibitem{Manchester+1995}
\bibinfo{author}{{Manchester}, R.~N.}
\newblock \bibinfo{title}{{The shape of pulsar beams.}}
\newblock \emph{\bibinfo{journal}{Journal of Astrophysics and Astronomy}}
  \textbf{\bibinfo{volume}{16}}, \bibinfo{pages}{107--117}
  (\bibinfo{year}{1995}).

\bibitem{Radhakrishnan69}
\bibinfo{author}{{Radhakrishnan}, V.} \& \bibinfo{author}{{Cooke}, D.~J.}
\newblock \bibinfo{title}{{Magnetic Poles and the Polarization Structure of
  Pulsar Radiation}}.
\newblock \emph{\bibinfo{journal}{Astrophys. Letter}}
  \textbf{\bibinfo{volume}{3}}, \bibinfo{pages}{225} (\bibinfo{year}{1969}).

\bibitem{rs75}
\bibinfo{author}{{Ruderman}, M.~A.} \& \bibinfo{author}{{Sutherland}, P.~G.}
\newblock \bibinfo{title}{{Theory of pulsars: polar gaps, sparks, and coherent
  microwave radiation.}}
\newblock \emph{\bibinfo{journal}{\apj}} \textbf{\bibinfo{volume}{196}},
  \bibinfo{pages}{51--72} (\bibinfo{year}{1975}).

\bibitem{okj20}
\bibinfo{author}{{Oswald}, L.}, \bibinfo{author}{{Karastergiou}, A.} \&
  \bibinfo{author}{{Johnston}, S.}
\newblock \bibinfo{title}{{Pulsar polarimetry with the Parkes ultra-wideband
  receiver}}.
\newblock \emph{\bibinfo{journal}{\mnras}} \textbf{\bibinfo{volume}{496}},
  \bibinfo{pages}{1418--1429} (\bibinfo{year}{2020}).

\bibitem{Backer1970}
\bibinfo{author}{{Backer}, D.~C.}
\newblock \bibinfo{title}{{Pulsar Nulling Phenomena}}.
\newblock \emph{\bibinfo{journal}{nature}} \textbf{\bibinfo{volume}{228}},
  \bibinfo{pages}{42--43} (\bibinfo{year}{1970}).

\bibitem{Ritchings1976}
\bibinfo{author}{{Ritchings}, R.~T.}
\newblock \bibinfo{title}{{Pulsar single pulse intensity measurements and pulse
  nulling.}}
\newblock \emph{\bibinfo{journal}{\mnras}} \textbf{\bibinfo{volume}{176}},
  \bibinfo{pages}{249--263} (\bibinfo{year}{1976}).

\bibitem{Sturrock1971}
\bibinfo{author}{{Sturrock}, P.~A.}
\newblock \bibinfo{title}{{A Model of Pulsars}}.
\newblock \emph{\bibinfo{journal}{\apj}} \textbf{\bibinfo{volume}{164}},
  \bibinfo{pages}{529} (\bibinfo{year}{1971}).

\bibitem{pts20}
\bibinfo{author}{{Philippov}, A.}, \bibinfo{author}{{Timokhin}, A.} \&
  \bibinfo{author}{{Spitkovsky}, A.}
\newblock \bibinfo{title}{{Origin of Pulsar Radio Emission}}.
\newblock \emph{\bibinfo{journal}{\prl}} \textbf{\bibinfo{volume}{124}},
  \bibinfo{pages}{245101} (\bibinfo{year}{2020}).

\bibitem{2020ApJ...889...69C}
\bibinfo{author}{{Chen}, A.~Y.}, \bibinfo{author}{{Cruz}, F.} \&
  \bibinfo{author}{{Spitkovsky}, A.}
\newblock \bibinfo{title}{{Filling the Magnetospheres of Weak Pulsars}}.
\newblock \emph{\bibinfo{journal}{\apj}} \textbf{\bibinfo{volume}{889}},
  \bibinfo{pages}{69} (\bibinfo{year}{2020}).

\bibitem{cgc+21}
\bibinfo{author}{{Cruz}, F.}, \bibinfo{author}{{Grismayer}, T.},
  \bibinfo{author}{{Chen}, A.~Y.}, \bibinfo{author}{{Spitkovsky}, A.} \&
  \bibinfo{author}{{Silva}, L.~O.}
\newblock \bibinfo{title}{{Coherent Emission from QED Cascades in Pulsar Polar
  Caps}}.
\newblock \emph{\bibinfo{journal}{\apjl}} \textbf{\bibinfo{volume}{919}},
  \bibinfo{pages}{L4} (\bibinfo{year}{2021}).

\bibitem{bba22}
\bibinfo{author}{{Bransgrove}, A.}, \bibinfo{author}{{Beloborodov}, A.~M.} \&
  \bibinfo{author}{{Levin}, Y.}
\newblock \bibinfo{title}{{Radio Emission and Electric Gaps in Pulsar
  Magnetospheres}}.
\newblock \emph{\bibinfo{journal}{arXiv e-prints}}
  \bibinfo{pages}{arXiv:2209.11362} (\bibinfo{year}{2022}).

\bibitem{klo+06}
\bibinfo{author}{{Kramer}, M.}, \bibinfo{author}{{Lyne}, A.~G.},
  \bibinfo{author}{{O'Brien}, J.~T.}, \bibinfo{author}{{Jordan}, C.~A.} \&
  \bibinfo{author}{{Lorimer}, D.~R.}
\newblock \bibinfo{title}{{A Periodically Active Pulsar Giving Insight into
  Magnetospheric Physics}}.
\newblock \emph{\bibinfo{journal}{Science}} \textbf{\bibinfo{volume}{312}},
  \bibinfo{pages}{549--551} (\bibinfo{year}{2006}).

\bibitem{llm+12}
\bibinfo{author}{{Lorimer}, D.~R.} \emph{et~al.}
\newblock \bibinfo{title}{{Radio and X-Ray Observations of the Intermittent
  Pulsar J1832+0029}}.
\newblock \emph{\bibinfo{journal}{\apj}} \textbf{\bibinfo{volume}{758}},
  \bibinfo{pages}{141} (\bibinfo{year}{2012}).

\bibitem{crc+12}
\bibinfo{author}{{Camilo}, F.}, \bibinfo{author}{{Ransom}, S.~M.},
  \bibinfo{author}{{Chatterjee}, S.}, \bibinfo{author}{{Johnston}, S.} \&
  \bibinfo{author}{{Demorest}, P.}
\newblock \bibinfo{title}{{PSR J1841-0500: A Radio Pulsar That Mostly is Not
  There}}.
\newblock \emph{\bibinfo{journal}{\apj}} \textbf{\bibinfo{volume}{746}},
  \bibinfo{pages}{63} (\bibinfo{year}{2012}).

\bibitem{hlk+04}
\bibinfo{author}{{Hobbs}, G.}, \bibinfo{author}{{Lyne}, A.~G.},
  \bibinfo{author}{{Kramer}, M.}, \bibinfo{author}{{Martin}, C.~E.} \&
  \bibinfo{author}{{Jordan}, C.}
\newblock \bibinfo{title}{{Long-term timing observations of 374 pulsars}}.
\newblock \emph{\bibinfo{journal}{\mnras}} \textbf{\bibinfo{volume}{353}},
  \bibinfo{pages}{1311--1344} (\bibinfo{year}{2004}).

\bibitem{Davies1970}
\bibinfo{author}{{Davies}, J.~G.} \& \bibinfo{author}{{Large}, M.~I.}
\newblock \bibinfo{title}{{A single-pulse search for pulsars}}.
\newblock \emph{\bibinfo{journal}{\mnras}} \textbf{\bibinfo{volume}{149}},
  \bibinfo{pages}{301} (\bibinfo{year}{1970}).

\bibitem{Mitra2004}
\bibinfo{author}{{Mitra}, D.} \& \bibinfo{author}{{Li}, X.~H.}
\newblock \bibinfo{title}{{Comparing geometrical and delay radio emission
  heights in pulsars}}.
\newblock \emph{\bibinfo{journal}{\aap}} \textbf{\bibinfo{volume}{421}},
  \bibinfo{pages}{215--228} (\bibinfo{year}{2004}).

\bibitem{Zhang2007}
\bibinfo{author}{{Zhang}, H.}, \bibinfo{author}{{Qiao}, G.~J.},
  \bibinfo{author}{{Han}, J.~L.}, \bibinfo{author}{{Lee}, K.~J.} \&
  \bibinfo{author}{{Wang}, H.~G.}
\newblock \bibinfo{title}{{PSR B2111+46: a test of the inverse Compton
  scattering model of radio emission}}.
\newblock \emph{\bibinfo{journal}{\aap}} \textbf{\bibinfo{volume}{465}},
  \bibinfo{pages}{525--531} (\bibinfo{year}{2007}).

\bibitem{Thomas2010}
\bibinfo{author}{{Thomas}, R.~M.~C.} \& \bibinfo{author}{{Gangadhara}, R.~T.}
\newblock \bibinfo{title}{{Absolute emission altitude of pulsars: PSRs
  B1839+09, B1916+14, and B2111+46}}.
\newblock \emph{\bibinfo{journal}{\aap}} \textbf{\bibinfo{volume}{515}},
  \bibinfo{pages}{A86} (\bibinfo{year}{2010}).

\bibitem{Gajjar2012}
\bibinfo{author}{{Gajjar}, V.}, \bibinfo{author}{{Joshi}, B.~C.} \&
  \bibinfo{author}{{Kramer}, M.}
\newblock \bibinfo{title}{{A survey of nulling pulsars using the Giant
  Meterwave Radio Telescope}}.
\newblock \emph{\bibinfo{journal}{\mnras}} \textbf{\bibinfo{volume}{424}},
  \bibinfo{pages}{1197--1205} (\bibinfo{year}{2012}).

\bibitem{Han2021}
\bibinfo{author}{{Han}, J.~L.} \emph{et~al.}
\newblock \bibinfo{title}{{The FAST Galactic Plane Pulsar Snapshot survey: I.
  Project design and pulsar discoveries}}.
\newblock \emph{\bibinfo{journal}{Research in Astronomy and Astrophysics}}
  \textbf{\bibinfo{volume}{21}}, \bibinfo{pages}{107} (\bibinfo{year}{2021}).

\bibitem{yws+14}
\bibinfo{author}{{Young}, N.~J.}, \bibinfo{author}{{Weltevrede}, P.},
  \bibinfo{author}{{Stappers}, B.~W.}, \bibinfo{author}{{Lyne}, A.~G.} \&
  \bibinfo{author}{{Kramer}, M.}
\newblock \bibinfo{title}{{On the apparent nulls and extreme variability of PSR
  J1107-5907}}.
\newblock \emph{\bibinfo{journal}{\mnras}} \textbf{\bibinfo{volume}{442}},
  \bibinfo{pages}{2519--2533} (\bibinfo{year}{2014}).

\bibitem{Burke2012}
\bibinfo{author}{{Burke-Spolaor}, S.} \emph{et~al.}
\newblock \bibinfo{title}{{The High Time Resolution Universe Pulsar Survey - V.
  Single-pulse energetics and modulation properties of 315 pulsars}}.
\newblock \emph{\bibinfo{journal}{\mnras}} \textbf{\bibinfo{volume}{423}},
  \bibinfo{pages}{1351--1367} (\bibinfo{year}{2012}).

\bibitem{cstt96}
\bibinfo{author}{{Cognard}, I.}, \bibinfo{author}{{Shrauner}, J.~A.},
  \bibinfo{author}{{Taylor}, J.~H.} \& \bibinfo{author}{{Thorsett}, S.~E.}
\newblock \bibinfo{title}{{Giant Radio Pulses from a Millisecond Pulsar}}.
\newblock \emph{\bibinfo{journal}{\apjl}} \textbf{\bibinfo{volume}{457}},
  \bibinfo{pages}{L81} (\bibinfo{year}{1996}).

\bibitem{rhc+1975}
\bibinfo{author}{{Rickett}, B.~J.}, \bibinfo{author}{{Hankins}, T.~H.} \&
  \bibinfo{author}{{Cordes}, J.~M.}
\newblock \bibinfo{title}{{The radio spectrum of micropulses from pulsar PSR
  0950+08.}}
\newblock \emph{\bibinfo{journal}{\apj}} \textbf{\bibinfo{volume}{201}},
  \bibinfo{pages}{425--430} (\bibinfo{year}{1975}).

\bibitem{spb+2004}
\bibinfo{author}{{Soglasnov}, V.~A.} \emph{et~al.}
\newblock \bibinfo{title}{{Giant Pulses from PSR B1937+21 with Widths $<=15$
  Nanoseconds and T$_{b}>=5{\times}10^{39}$ K, the Highest Brightness
  Temperature Observed in the Universe}}.
\newblock \emph{\bibinfo{journal}{\apj}} \textbf{\bibinfo{volume}{616}},
  \bibinfo{pages}{439--451} (\bibinfo{year}{2004}).

\bibitem{cr79}
\bibinfo{author}{{Cheng}, A.~F.} \& \bibinfo{author}{{Ruderman}, M.~A.}
\newblock \bibinfo{title}{{A theory of subpulse polarization patterns from
  radio pulsars.}}
\newblock \emph{\bibinfo{journal}{\apj}} \textbf{\bibinfo{volume}{229}},
  \bibinfo{pages}{348--360} (\bibinfo{year}{1979}).

\bibitem{wlh10}
\bibinfo{author}{{Wang}, C.}, \bibinfo{author}{{Lai}, D.} \&
  \bibinfo{author}{{Han}, J.}
\newblock \bibinfo{title}{{Polarization changes of pulsars due to wave
  propagation through magnetospheres}}.
\newblock \emph{\bibinfo{journal}{\mnras}} \textbf{\bibinfo{volume}{403}},
  \bibinfo{pages}{569--588} (\bibinfo{year}{2010}).

\bibitem{bp12}
\bibinfo{author}{{Beskin}, V.~S.} \& \bibinfo{author}{{Philippov}, A.~A.}
\newblock \bibinfo{title}{{On the mean profiles of radio pulsars - I. Theory of
  propagation effects}}.
\newblock \emph{\bibinfo{journal}{\mnras}} \textbf{\bibinfo{volume}{425}},
  \bibinfo{pages}{814--840} (\bibinfo{year}{2012}).

\bibitem{zr05}
\bibinfo{author}{{Srostlik}, Z.} \& \bibinfo{author}{{Rankin}, J.~M.}
\newblock \bibinfo{title}{{Core and conal component analysis of pulsar
  B1237+25}}.
\newblock \emph{\bibinfo{journal}{\mnras}} \textbf{\bibinfo{volume}{362}},
  \bibinfo{pages}{1121--1133} (\bibinfo{year}{2005}).

\bibitem{yws+15}
\bibinfo{author}{{Young}, N.~J.}, \bibinfo{author}{{Weltevrede}, P.},
  \bibinfo{author}{{Stappers}, B.~W.}, \bibinfo{author}{{Lyne}, A.~G.} \&
  \bibinfo{author}{{Kramer}, M.}
\newblock \bibinfo{title}{{Long-term observations of three nulling pulsars}}.
\newblock \emph{\bibinfo{journal}{\mnras}} \textbf{\bibinfo{volume}{449}},
  \bibinfo{pages}{1495--1504} (\bibinfo{year}{2015}).

\bibitem{mhth05}
\bibinfo{author}{{Manchester}, R.~N.}, \bibinfo{author}{{Hobbs}, G.~B.},
  \bibinfo{author}{{Teoh}, A.} \& \bibinfo{author}{{Hobbs}, M.}
\newblock \bibinfo{title}{{The Australia Telescope National Facility Pulsar
  Catalogue}}.
\newblock \emph{\bibinfo{journal}{\aj}} \textbf{\bibinfo{volume}{129}},
  \bibinfo{pages}{1993--2006} (\bibinfo{year}{2005}).

\bibitem{dspsr}
\bibinfo{author}{van Straten, W.} \& \bibinfo{author}{Bailes, M.}
\newblock \bibinfo{title}{Dspsr: Digital signal processing software for pulsar
  astronomy}.
\newblock \emph{\bibinfo{journal}{pasa}} \textbf{\bibinfo{volume}{28}},
  \bibinfo{pages}{1--14} (\bibinfo{year}{2011}).

\bibitem{whx+23}
\bibinfo{author}{{Wang}, P.~F.} \emph{et~al.}
\newblock \bibinfo{title}{{FAST pulsar database: I. polarization profiles of
  644 pulsars}}.
\newblock \emph{\bibinfo{journal}{RAA}} \textbf{\bibinfo{volume}{Submitted}}
  (\bibinfo{year}{2023}).

\bibitem{psrchive}
\bibinfo{author}{Hotan, A.~W.}, \bibinfo{author}{van Straten, W.} \&
  \bibinfo{author}{Manchester, R.~N.}
\newblock \bibinfo{title}{Psrchive and psrfits: An open approach to radio
  pulsar data storage and analysis}.
\newblock \emph{\bibinfo{journal}{pasa}} \textbf{\bibinfo{volume}{21}},
  \bibinfo{pages}{302--309} (\bibinfo{year}{2004}).

\bibitem{fdr15}
\bibinfo{author}{{Force}, M.~M.}, \bibinfo{author}{{Demorest}, P.} \&
  \bibinfo{author}{{Rankin}, J.~M.}
\newblock \bibinfo{title}{{Absolute polarization determinations of 33 pulsars
  using the Green Bank Telescope}}.
\newblock \emph{\bibinfo{journal}{\mnras}} \textbf{\bibinfo{volume}{453}},
  \bibinfo{pages}{4485--4499} (\bibinfo{year}{2015}).

\bibitem{Lyne1988}
\bibinfo{author}{{Lyne}, A.~G.} \& \bibinfo{author}{{Manchester}, R.~N.}
\newblock \bibinfo{title}{{The shape of pulsar radio beams.}}
\newblock \emph{\bibinfo{journal}{\mnras}} \textbf{\bibinfo{volume}{234}},
  \bibinfo{pages}{477--508} (\bibinfo{year}{1988}).

\bibitem{Gould1998mn}
\bibinfo{author}{{Gould}, D.~M.} \& \bibinfo{author}{{Lyne}, A.~G.}
\newblock \bibinfo{title}{{Multifrequency polarimetry of 300 radio pulsars}}.
\newblock \emph{\bibinfo{journal}{\mnras}} \textbf{\bibinfo{volume}{301}},
  \bibinfo{pages}{235--260} (\bibinfo{year}{1998}).

\bibitem{Force2015}
\bibinfo{author}{{Force}, M.~M.}, \bibinfo{author}{{Demorest}, P.} \&
  \bibinfo{author}{{Rankin}, J.~M.}
\newblock \bibinfo{title}{{Absolute polarization determinations of 33 pulsars
  using the Green Bank Telescope}}.
\newblock \emph{\bibinfo{journal}{\mnras}} \textbf{\bibinfo{volume}{453}},
  \bibinfo{pages}{4485--4499} (\bibinfo{year}{2015}).

\bibitem{ran93}
\bibinfo{author}{{Rankin}, J.~M.}
\newblock \bibinfo{title}{{Toward an Empirical Theory of Pulsar Emission. VI.
  The Geometry of the Conal Emission Region}}.
\newblock \emph{\bibinfo{journal}{\apj}} \textbf{\bibinfo{volume}{405}},
  \bibinfo{pages}{285} (\bibinfo{year}{1993}).

\bibitem{rr90}
\bibinfo{author}{{Radhakrishnan}, V.} \& \bibinfo{author}{{Rankin}, J.~M.}
\newblock \bibinfo{title}{{Toward an Empirical Theory of Pulsar Emission. V. On
  the Circular Polarization in Pulsar Radiation}}.
\newblock \emph{\bibinfo{journal}{\apj}} \textbf{\bibinfo{volume}{352}},
  \bibinfo{pages}{258} (\bibinfo{year}{1990}).

\bibitem{hmxq98}
\bibinfo{author}{{Han}, J.~L.}, \bibinfo{author}{{Manchester}, R.~N.},
  \bibinfo{author}{{Xu}, R.~X.} \& \bibinfo{author}{{Qiao}, G.~J.}
\newblock \bibinfo{title}{{Circular polarization in pulsar integrated
  profiles}}.
\newblock \emph{\bibinfo{journal}{\mnras}} \textbf{\bibinfo{volume}{300}},
  \bibinfo{pages}{373--387} (\bibinfo{year}{1998}).

\bibitem{ghw21}
\bibinfo{author}{{Gangadhara}, R.~T.}, \bibinfo{author}{{Han}, J.~L.} \&
  \bibinfo{author}{{Wang}, P.~F.}
\newblock \bibinfo{title}{{Coherent Curvature Radio Emission and Polarization
  from Pulsars}}.
\newblock \emph{\bibinfo{journal}{\apj}} \textbf{\bibinfo{volume}{911}},
  \bibinfo{pages}{152} (\bibinfo{year}{2021}).

\bibitem{1993ApJ...402..264C}
\bibinfo{author}{{Chen}, K.} \& \bibinfo{author}{{Ruderman}, M.}
\newblock \bibinfo{title}{{Pulsar Death Lines and Death Valley}}.
\newblock \emph{\bibinfo{journal}{\apj}} \textbf{\bibinfo{volume}{402}},
  \bibinfo{pages}{264} (\bibinfo{year}{1993}).

\bibitem{zhm00}
\bibinfo{author}{{Zhang}, B.}, \bibinfo{author}{{Harding}, A.~K.} \&
  \bibinfo{author}{{Muslimov}, A.~G.}
\newblock \bibinfo{title}{{Radio Pulsar Death Line Revisited: Is PSR J2144-3933
  Anomalous?}}
\newblock \emph{\bibinfo{journal}{\apjl}} \textbf{\bibinfo{volume}{531}},
  \bibinfo{pages}{L135--L138} (\bibinfo{year}{2000}).

\bibitem{wwsr06}
\bibinfo{author}{{Weltevrede}, P.}, \bibinfo{author}{{Wright}, G.~A.~E.},
  \bibinfo{author}{{Stappers}, B.~W.} \& \bibinfo{author}{{Rankin}, J.~M.}
\newblock \bibinfo{title}{{The bright spiky emission of pulsar B0656+14}}.
\newblock \emph{\bibinfo{journal}{\aap}} \textbf{\bibinfo{volume}{458}},
  \bibinfo{pages}{269--283} (\bibinfo{year}{2006}).

\bibitem{bb10}
\bibinfo{author}{{Burke-Spolaor}, S.} \& \bibinfo{author}{{Bailes}, M.}
\newblock \bibinfo{title}{{The millisecond radio sky: transients from a blind
  single-pulse search}}.
\newblock \emph{\bibinfo{journal}{\mnras}} \textbf{\bibinfo{volume}{402}},
  \bibinfo{pages}{855--866} (\bibinfo{year}{2010}).

\bibitem{eamn12}
\bibinfo{author}{{Esamdin}, A.}, \bibinfo{author}{{Abdurixit}, D.},
  \bibinfo{author}{{Manchester}, R.~N.} \& \bibinfo{author}{{Niu}, H.~B.}
\newblock \bibinfo{title}{{PSR B0826-34: Sometimes a Rotating Radio
  Transient}}.
\newblock \emph{\bibinfo{journal}{\apjl}} \textbf{\bibinfo{volume}{759}},
  \bibinfo{pages}{L3} (\bibinfo{year}{2012}).

\bibitem{gj69}
\bibinfo{author}{{Goldreich}, P.} \& \bibinfo{author}{{Julian}, W.~H.}
\newblock \bibinfo{title}{{Pulsar Electrodynamics}}.
\newblock \emph{\bibinfo{journal}{\apj}} \textbf{\bibinfo{volume}{157}},
  \bibinfo{pages}{869} (\bibinfo{year}{1969}).

\bibitem{as79}
\bibinfo{author}{{Arons}, J.} \& \bibinfo{author}{{Scharlemann}, E.~T.}
\newblock \bibinfo{title}{{Pair formation above pulsar polar caps: structure of
  the low altitude acceleration zone.}}
\newblock \emph{\bibinfo{journal}{\apj}} \textbf{\bibinfo{volume}{231}},
  \bibinfo{pages}{854--879} (\bibinfo{year}{1979}).

\bibitem{chr86}
\bibinfo{author}{{Cheng}, K.~S.}, \bibinfo{author}{{Ho}, C.} \&
  \bibinfo{author}{{Ruderman}, M.}
\newblock \bibinfo{title}{{Energetic Radiation from Rapidly Spinning Pulsars.
  I. Outer Magnetosphere Gaps}}.
\newblock \emph{\bibinfo{journal}{\apj}} \textbf{\bibinfo{volume}{300}},
  \bibinfo{pages}{500} (\bibinfo{year}{1986}).

\bibitem{qlw+04}
\bibinfo{author}{{Qiao}, G.~J.}, \bibinfo{author}{{Lee}, K.~J.},
  \bibinfo{author}{{Wang}, H.~G.}, \bibinfo{author}{{Xu}, R.~X.} \&
  \bibinfo{author}{{Han}, J.~L.}
\newblock \bibinfo{title}{{The Inner Annular Gap for Pulsar Radiation:
  {\ensuremath{\gamma}}-Ray and Radio Emission}}.
\newblock \emph{\bibinfo{journal}{\apjl}} \textbf{\bibinfo{volume}{606}},
  \bibinfo{pages}{L49--L52} (\bibinfo{year}{2004}).

\bibitem{mh04}
\bibinfo{author}{{Muslimov}, A.~G.} \& \bibinfo{author}{{Harding}, A.~K.}
\newblock \bibinfo{title}{{High-Altitude Particle Acceleration and Radiation in
  Pulsar Slot Gaps}}.
\newblock \emph{\bibinfo{journal}{\apj}} \textbf{\bibinfo{volume}{606}},
  \bibinfo{pages}{1143--1153} (\bibinfo{year}{2004}).

\end{thebibliography}

\clearpage

\begin{figure*}
    \centering
    \includegraphics[width=\columnwidth]{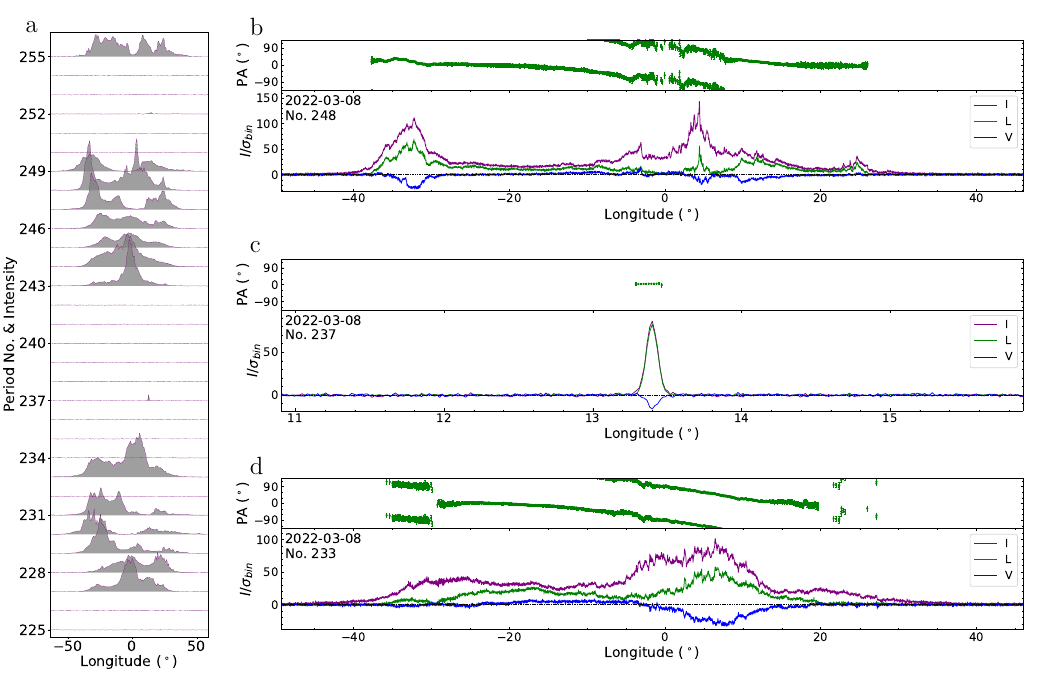}
    \caption{Fig.1 $|$ FAST detection of a dwarf pulse in a series pulses of B2111+46. (a) A segment of pulse trains of PSR B2111+46 observed in the session of 2022-03-08 by the FAST, showing some emission and nulling periods; (b-d): polarization profiles of 3 individual pulses. In the lower sub-panels the total intensity $I$, linear polarization $L$ and circular polarization $V$ (with positive values for the left hand sense) are plotted in the original time resolution (49.152~$\mu$s) of FAST observations, and the polarization angles (PA) are plotted in the upper sub-panels. The dwarf pulse in the period No.237 has only one resolved emission cell, almost fully linearly polarized with a width of about $0.1^{\circ}$. Many notches of the other two pulse profiles are sensitive significant detection of real intensity fluctuations caused by emission cells with different strength. The error bar for PA is  $\pm1\sigma$. The intensity is scaled with the off-pulse fluctuations expressed by $\sigma_{\rm bin}$.}
\end{figure*}

\begin{figure*}
    \centering
    \includegraphics[width=\columnwidth]{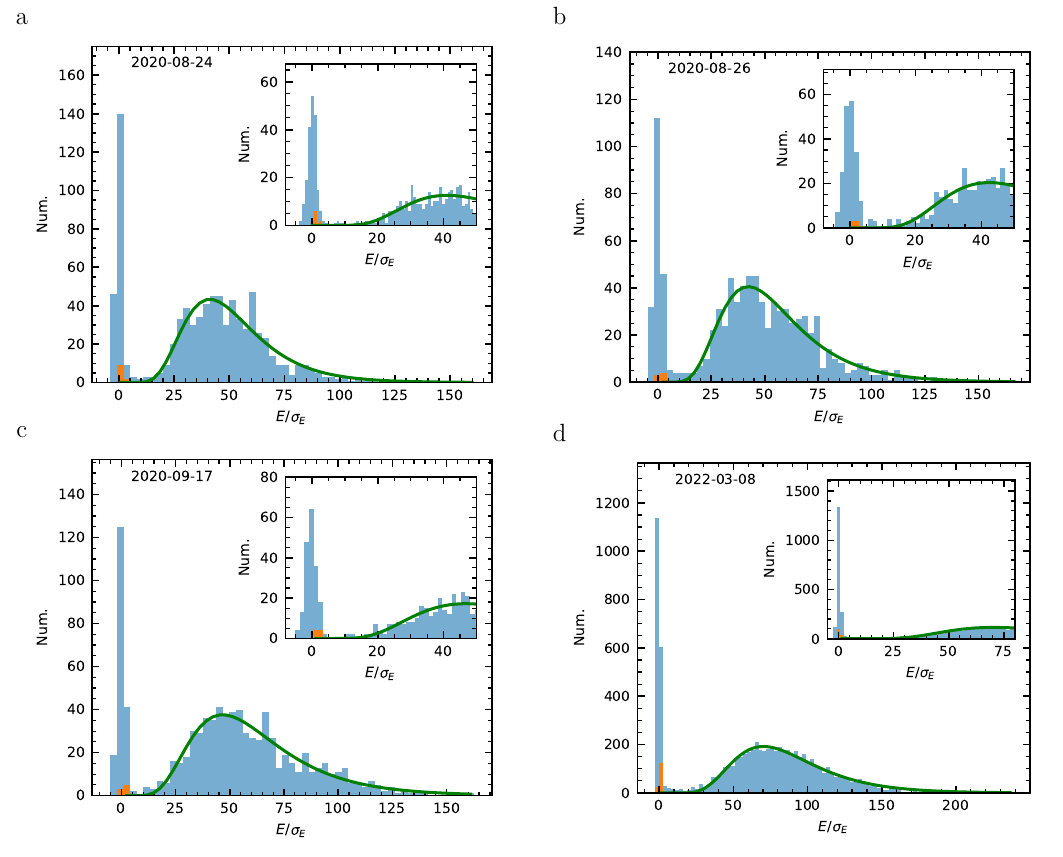}
    \caption{Fig.2 $|$ Dwarf pulses detected in the nulling periods with very low energy. The energy distribution of 822, 886, 885 and 7097 individual pulses of PSR B2111+46 observed in four FAST observation sessions. The pulse energy $E$ (i.e. the fluence) for every period is the sum of energy of an individual pulse over the full pulse-on window defined by the mean profile. To express data quality of the observations, the distribution is scaled by the standard deviations $\sigma_E$ of the stochastic energy in the same size but pulse-off window, rather than the averaged energy $\langle E \rangle$ as in literature that is too high for FAST detected dwarf pulses. The emission state and the null state show two main peaks in the histogram. The orange part overlapping on the nulling state stands for dwarf pulses. The green curve is the best log-normal fitting for the normal emission energy distribution.}
\end{figure*}

\begin{figure*}
    \centering
    \includegraphics[width=\columnwidth]{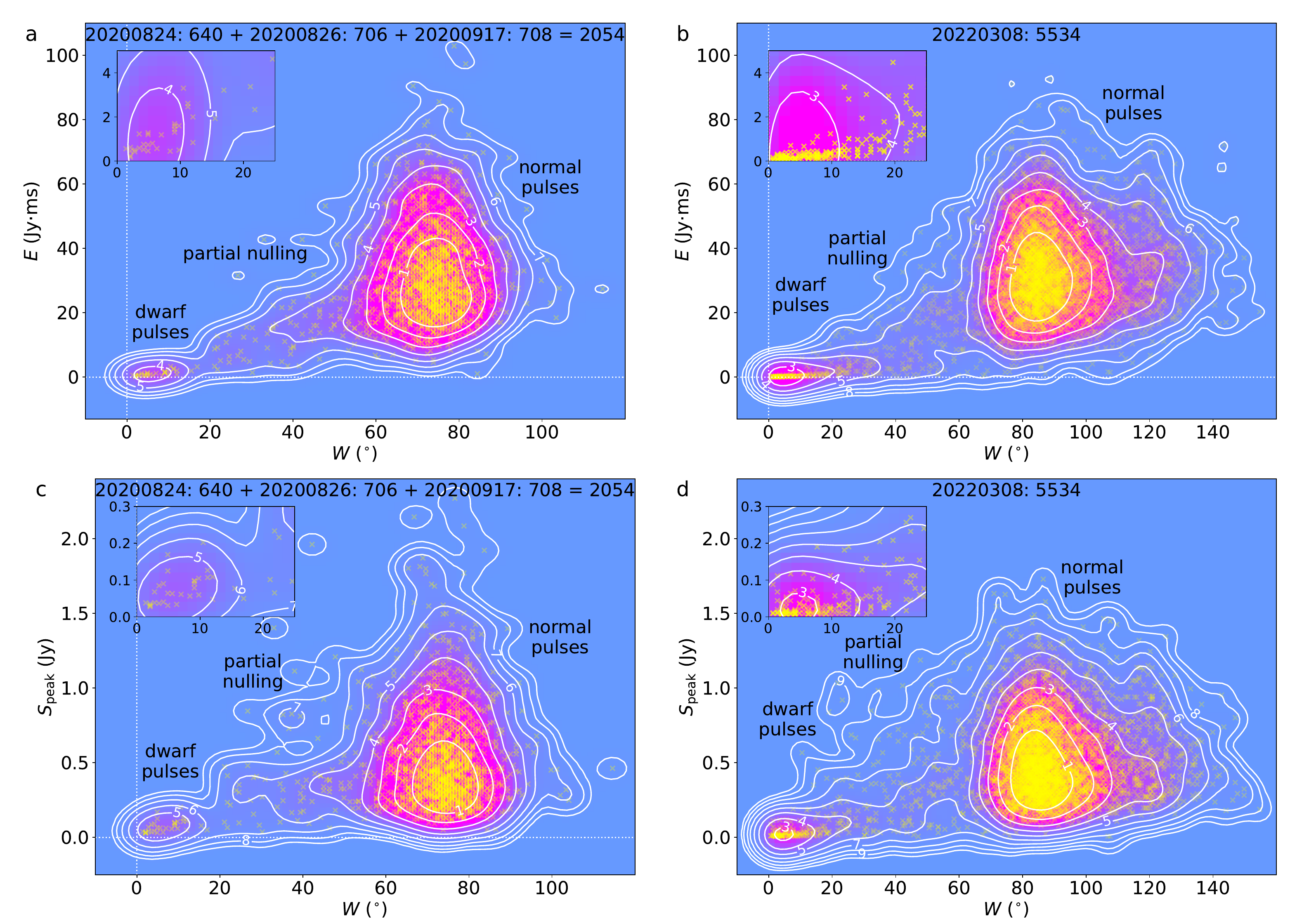}
    \caption{Fig.3 $|$ Dwarf pulses of PSR B2111+46 as a distinct population from the partial nulling and normal pulses. Pulse width is measured at the most outer profile at the 3$\sigma$ detection level. {\it Upper panels}:  Pulse fluence integrated over each pulse, $E$, against pulse width, $W$, and the density distribution of data are presented in color and also in contours at levels of 1/$2^{-n}$ of the maximum density ($n$ = 1 -- 8). More sensitive observations on 2022-03-08 gives larger widths for normal pulses. No mention of about 20\% of nulling period, normal pulses are concentrated around the main peak, with a fluence in the range from 10 to about 50~Jy$\cdot$ms and a pulse width $60^{\circ}<W<100^{\circ}$. The dwarf pulses are concentrated on another peak, with a fluence less than 1 Jy$\cdot$ms and a pulse width less than $15^{\circ}$ (i.e. 40 ms). In between are partially nulling pulses (see Method: Individual pulses). {\it Lower panels}: The same as {\it the upper panel} but for peak flux density $S$ against the pulse width. The {\it left panels} are made for individual pulses obtained the three sessions in 2020, the {\it right panels} for pulses detected in the longer verification observation session on 2022-03-08.}
\end{figure*}

\begin{figure*}
    \centering
    \includegraphics[width=\columnwidth]{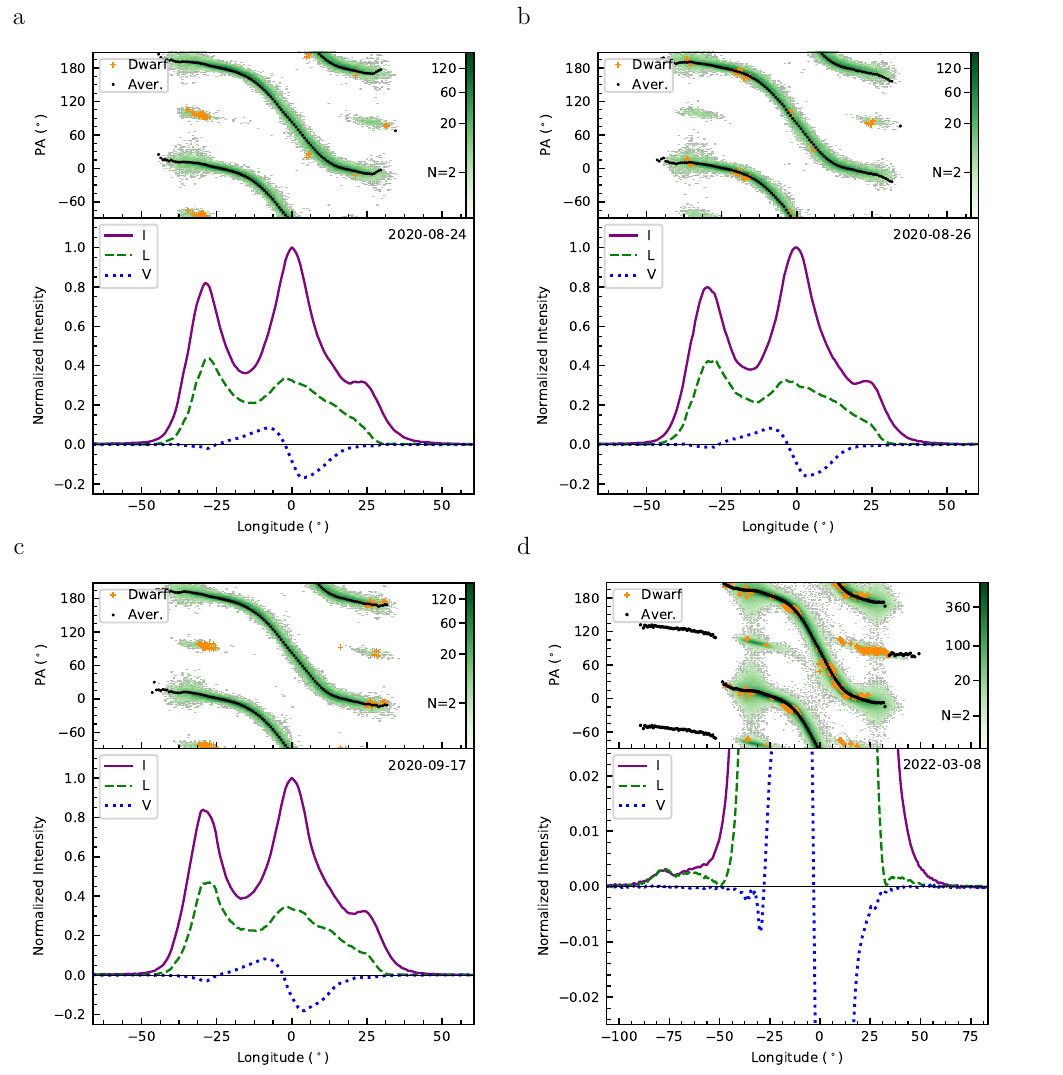}
    \caption{Fig.4 $|$ Polarization angle distribution of dwarf pulses compared to the data of normal pulses. The PA data of every bin of dwarf pulses (yellow) are plotted against those of normal pulses (green) and the mean polarization profiles in the 4 FAST observation sessions. The orthogonal modes are mostly predominant in  the wings of conal components. The magnified polarization profiles of 2022-03-08 illuminate the newly detected much extended leading weak profile wing. The error bar for PA is $\pm1\sigma$. The intensity is scaled with the peak value.}
\end{figure*}

\begin{figure*}
    \centering
    \includegraphics[width=\columnwidth]{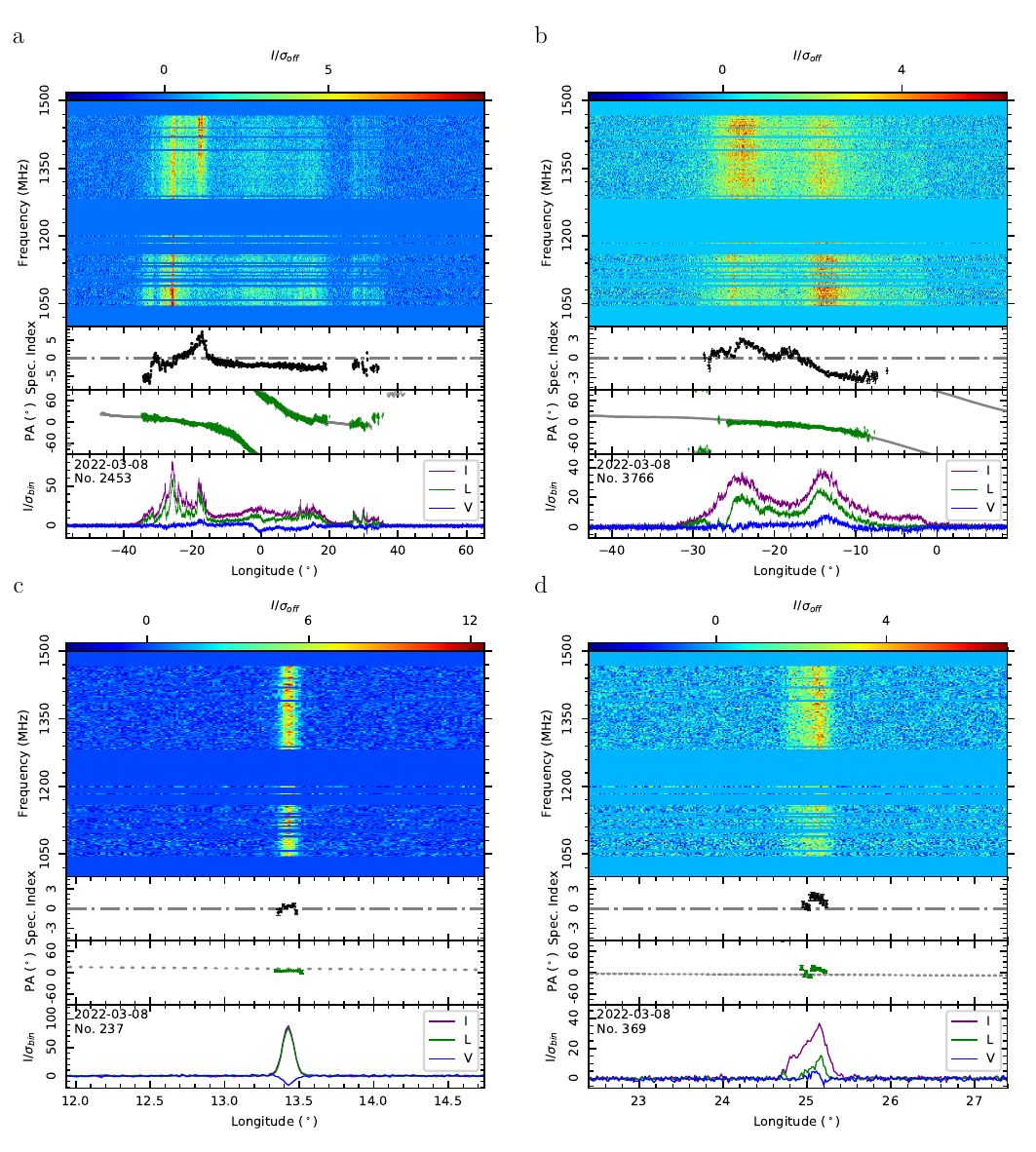}
    \caption{Fig.5 $|$ Various phase-resolved spectral index for two individual pulses and two dwarf pulses. The water-fall plot for the individual pulse intensity on the phase-frequency plane (upper subpanels. The frequency channels containing RFIs have been removed), clearly demonstrating the variation of phase-resolved spectral index for individual pulses (the second upper subpanels). The polarization profiles of the pulse and the PA values (green) together with the mean PA curve (grey) are plotted in the bottom subpanels and the second bottom subpanels. 
    The observation date and the period number of the individual pulse are marked in the bottom panel. The PA curves are fitted with the rotating vector model\cite{Radhakrishnan69}. The error bar for PA is  $\pm1\sigma$. The intensity is scaled with the off-pulse fluctuations expressed by $\sigma_{\rm bin}$.}
\end{figure*}

\begin{figure*}
    \centering
    \includegraphics[width=\columnwidth]{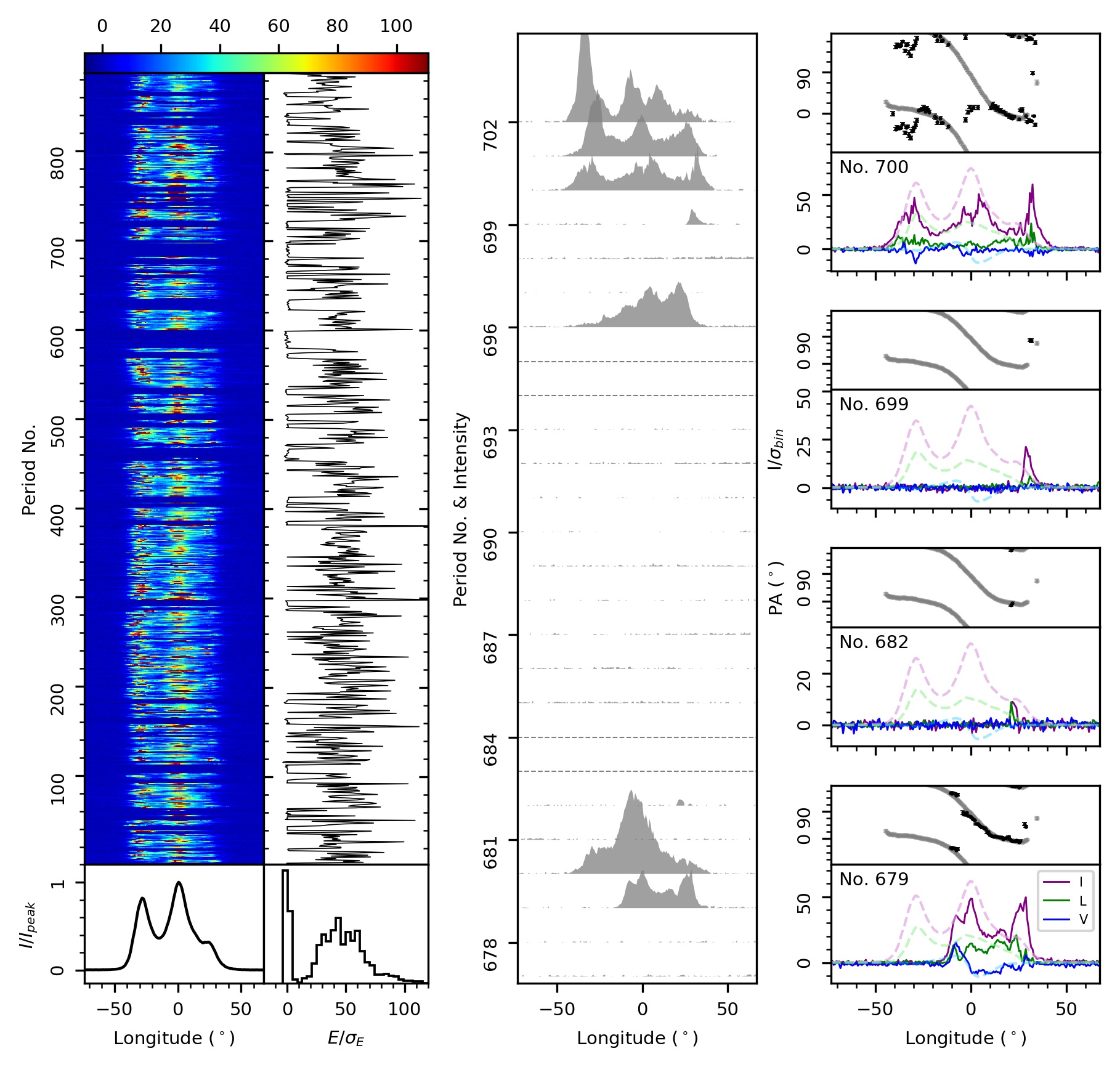}
    \caption{Extended Data Fig.1 $|$ The pulses of PSR B2111+46 observed by the FAST on the session on 2020-08-24. The most left panel is the train of individual pulses for 886 periods, with the mean profile shown in the bottom, and the intensity is normalized using the peak value. The total energy of every individual pulse is plotted in the immediately right, so that the energy fluctuations are seen very clearly which show the two predominate peaks for both nulling and emission states in the number distributions in the bottom. A segment of the pulse stack is shown in grey for high quality individual pulses, with significant fluctuations of profile amplitude, in which normal individual pulses {can be seen} in the period No. 702-700, 696 and 680, partial nulling in the period No. 679, and dwarf pulses of the period No. 699 and 682. The detailed polarization profiles for 4 pulses are presented in the right 4 panels, each with total intensity $I$, linear polarization $L$ and circular polarization $V$ in the bottom subpanel and $PA$ in the upper subpanel. The polarization profiles of the mean pulse are shown in dashed line in these subpanels for comparison. The error bar for PA is  $\pm1\sigma$. }
\end{figure*}

\begin{figure*}
    \centering
    \includegraphics[width=\columnwidth]{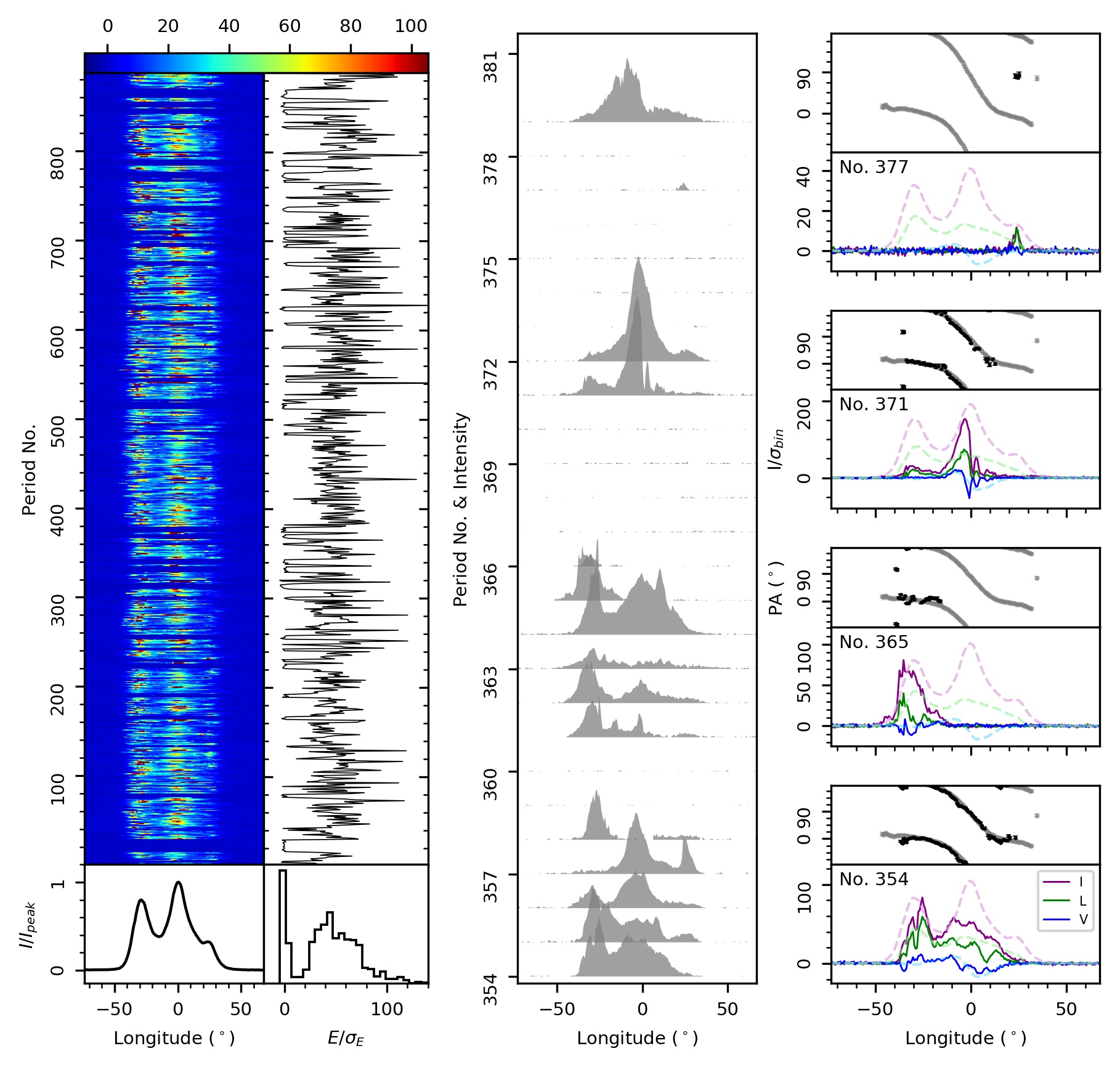}
    \caption{Extended Data Fig.2 $|$ The pulses of PSR B2111+46 observed by the FAST on the session on  2020-08-26. The most left panel is the train of individual pulses for 886 periods, with the mean profile shown in the bottom, and the intensity is normalized using the peak value. The total energy of every individual pulse is plotted in the immediately right. A segment of the pulse stack is shown in grey for high quality individual pulses, with a dwarf pulse in the period No. 377 and partial nulling in the period No. 365. The detailed polarization profiles for 4 pulses are presented in the right 4 panels, each with total intensity $I$, linear polarization $L$ and circular polarization $V$ in the bottom subpanel and $PA$ in the upper subpanel. The polarization profiles of the mean pulse are shown in dashed line in these subpanels for comparison. The error bar for PA is  $\pm1\sigma$.}
\end{figure*}

\begin{figure*}
    \centering
    \includegraphics[width=\columnwidth]{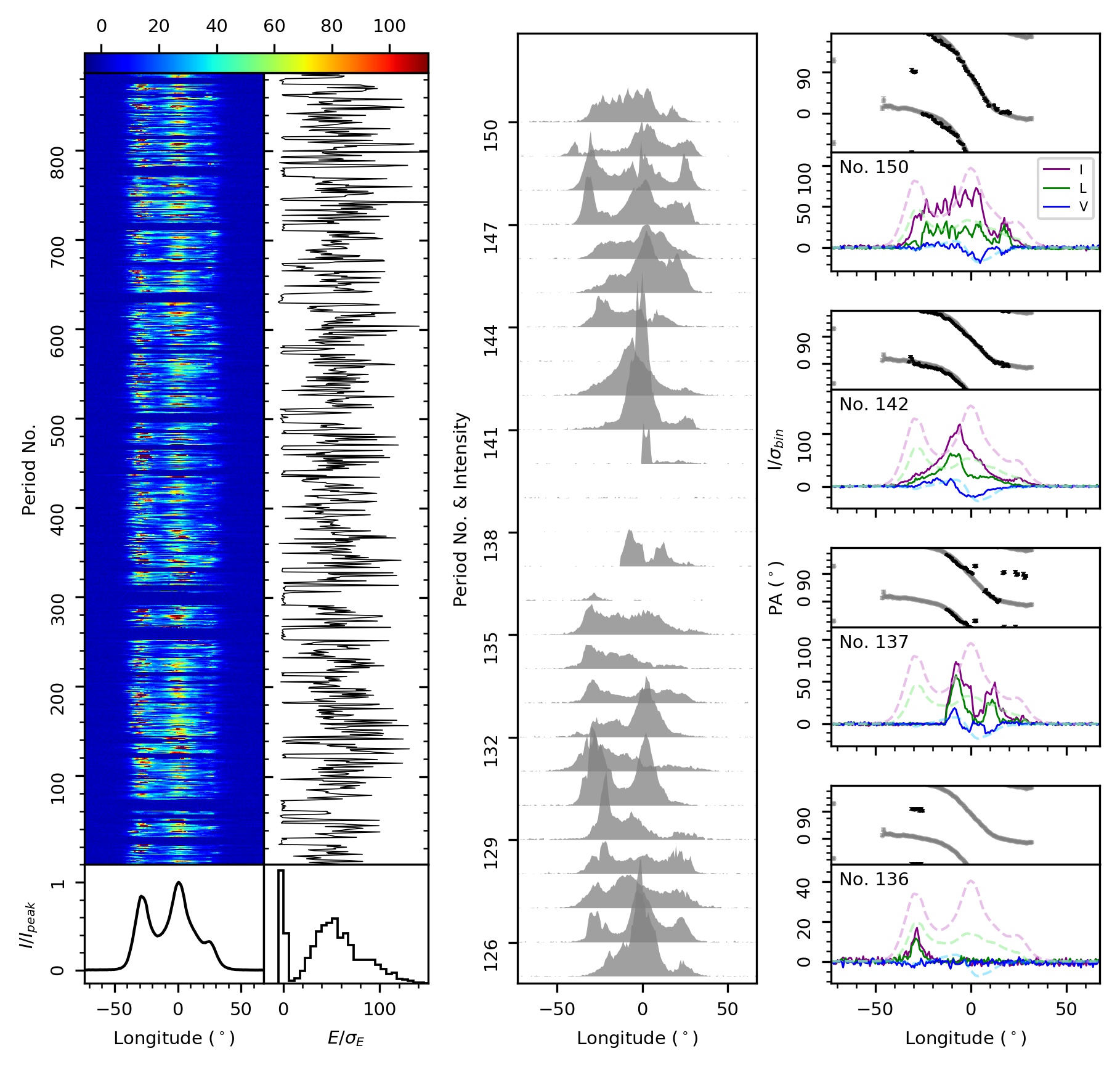}
    \caption{Extended Data Fig.3 $|$ The pulses of PSR B2111+46 observed by the FAST on the session on 2020-09-17.
    The most left panel is the train of individual pulses for 885 periods, with the mean profile shown in the bottom, and the intensity is normalized using the peak value. The total energy of every individual pulse is plotted in the immediately right. A segment of the pulse stack is shown in grey for high quality individual pulses, with a dwarf pulse in the period No. 136 and partial nulling in the period No. 137. The detailed polarization profiles for 4 pulses are presented in the right 4 panels, each with total intensity $I$, linear polarization $L$ and circular polarization $V$ in the bottom subpanel and $PA$ in the upper subpanel. The polarization profiles of the mean pulse are shown in dashed line in these subpanels for comparison. The error bar for PA is  $\pm1\sigma$.
    }
\end{figure*}

\begin{figure*}
    \centering
    \includegraphics[width=\columnwidth]{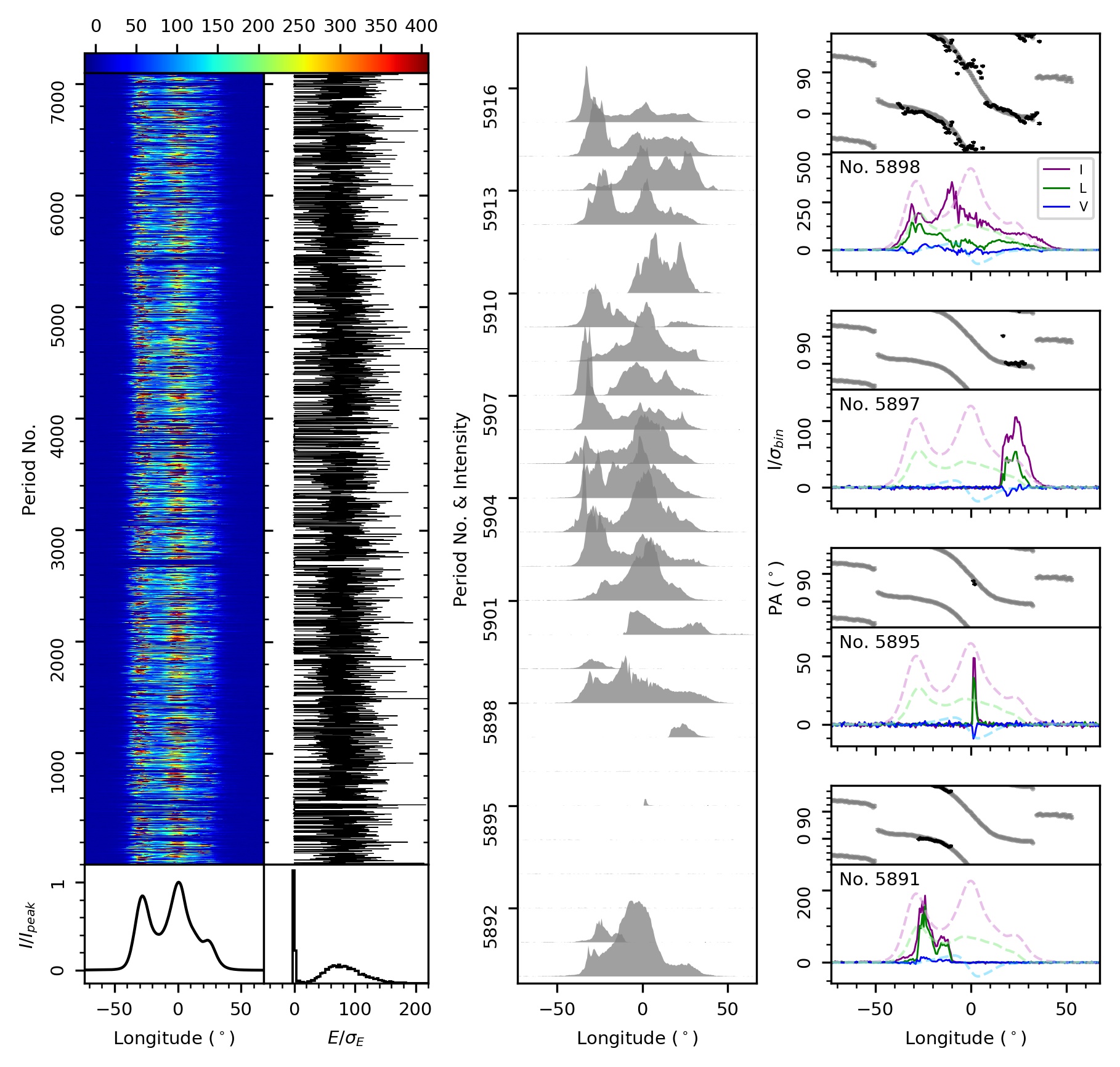}
    \caption{Extended Data Fig.4 $|$ The pulses of PSR B2111+46 observed by the FAST on the session on on 2022-03-08.
    The most left panel is the train of individual pulses for 7098 periods, with the mean profile shown in the bottom, and the intensity is normalized using the peak value. The total energy of every individual pulse is plotted in the immediately right. A segment of the pulse stack is shown in grey for high quality individual pulses, with a dwarf pulse in period of No.5895 and two partial nullings in the period No. 5891 and 5897. The detailed polarization profiles for 4 pulses are presented in the right 4 panels, each with total intensity $I$, linear polarization $L$ and circular polarization $V$ in the bottom subpanel and $PA$ in the upper subpanel. The polarization profiles of the mean pulse are shown in dashed line in these subpanels for comparison. The error bar for PA is  $\pm1\sigma$.
    }
\end{figure*}

\begin{figure*}
    \centering
    \includegraphics[width=0.9\columnwidth]{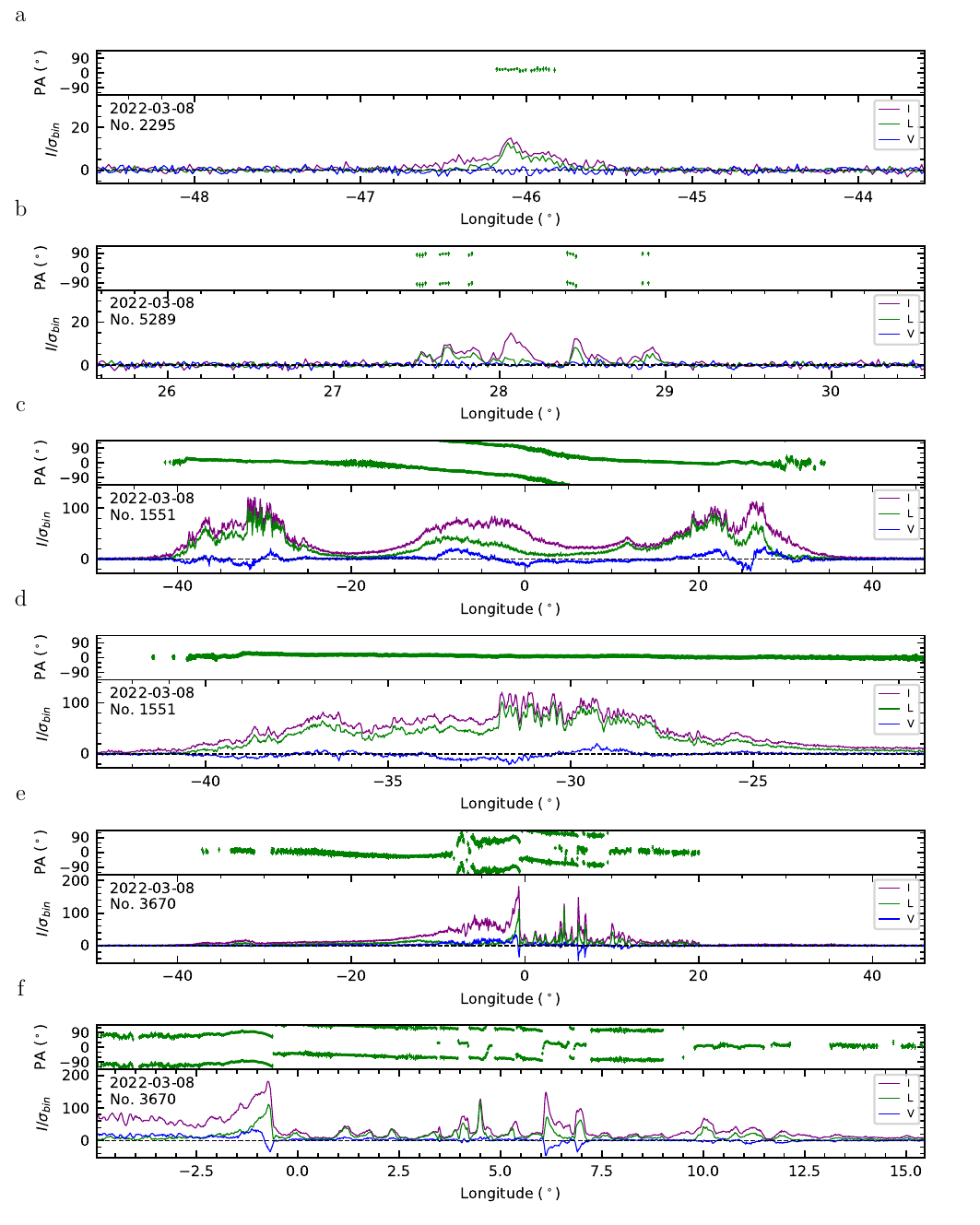}
    \caption{Extended Data Fig.5 $|$ Examples of polarization profiles for two dwarf pulses and two strong individual pulses in high time resolution. All of them are  observed on 2022-03-08 by FAST with time resolution of 49.152~$\mu$s. Polarization profiles for two strong individual pulses are shown for their elongated central part in the next panel. Each rip in the profiles is real, well-significant above the noise fluctuations. These unrepresented details indicate that the observed individual pulses are an incoherent collection of many elementary pulses generated {separately} in the magnetosphere. The error bar for PA is  $\pm1\sigma$. The intensity is scaled with the off-pulse fluctuations expressed by $\sigma_{\rm bin}$.}
\end{figure*}

\begin{figure*}
    \centering
    \includegraphics[width=0.8\columnwidth]{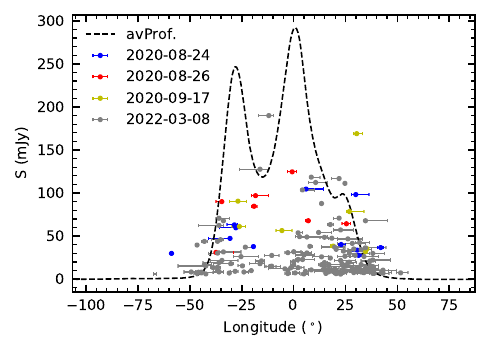}
    \caption{Extended Data Fig.6 $|$ Longitude distribution of dwarf pulses. Longitude distribution of dwarf pulses locations are compared to the mean pulse profile indicated by the dash line. 
The bar length stands for dwarf pulse width, and  the dots mark the peak locations in the longitude.}
\end{figure*}

\begin{figure*}
    \centering
    \includegraphics[width=0.8\columnwidth]{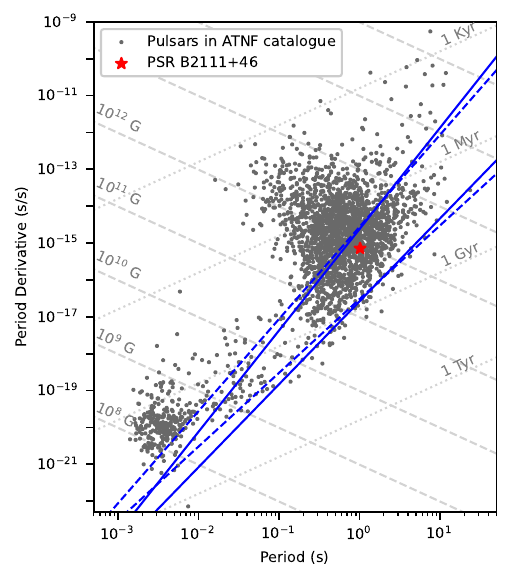}
    \caption{Extended Data Fig.7 $|$ Pulsar $P-\dot{P}$ diagram and the location of PSR B2111+46 in the death valley. The death lines are given for the curvature radiation in a dipole field (upper one) and an extremely curved field (lower one) in the vacuum gap model (sold lines) and the space-charged-limited flow model (dashed lines) given in \cite{zhm00}. All pulsar data are taken from the ATNF pulsar Catalogue\cite{mhth05} (version 1.70). The background gray dashed and dotted lines stand for constant surface magnetic field strengths and characteristic ages, respectively.}
\end{figure*}

\begin{figure*}
    \centering
    \includegraphics[width=8.8cm]{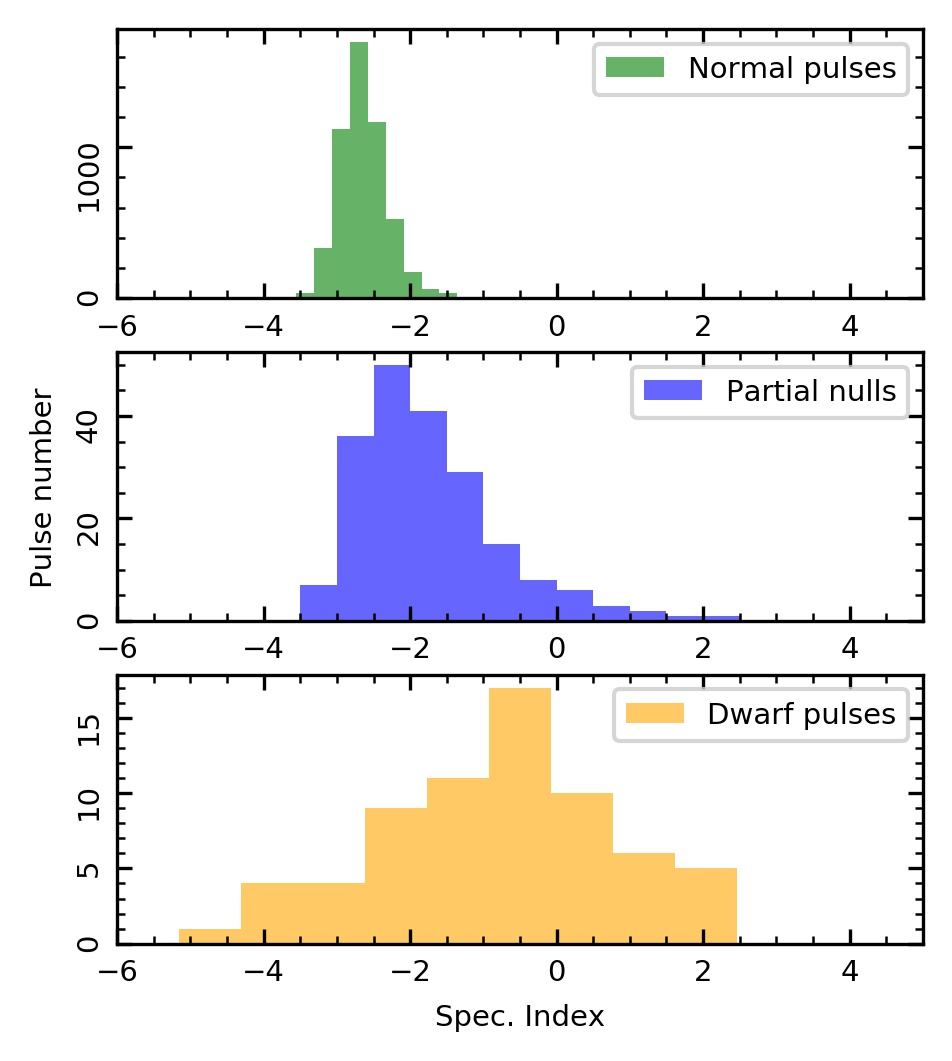}
    \caption{Extended Data Fig.8 $|$ Distributions for spectral indexes of three kinds of individual pulses. All these pulses, including 5175 normal pulses, 199 partial nulling pulses and 67 dwarf pulses, are observed by FAST on 2022-03-08. The indexes are calculated for each individual pulse by using the on-pulse integrated intensity, and have an uncertainty less than 0.5.}
\end{figure*}

\begin{figure*}
    \centering
   \includegraphics[width=0.8\columnwidth]{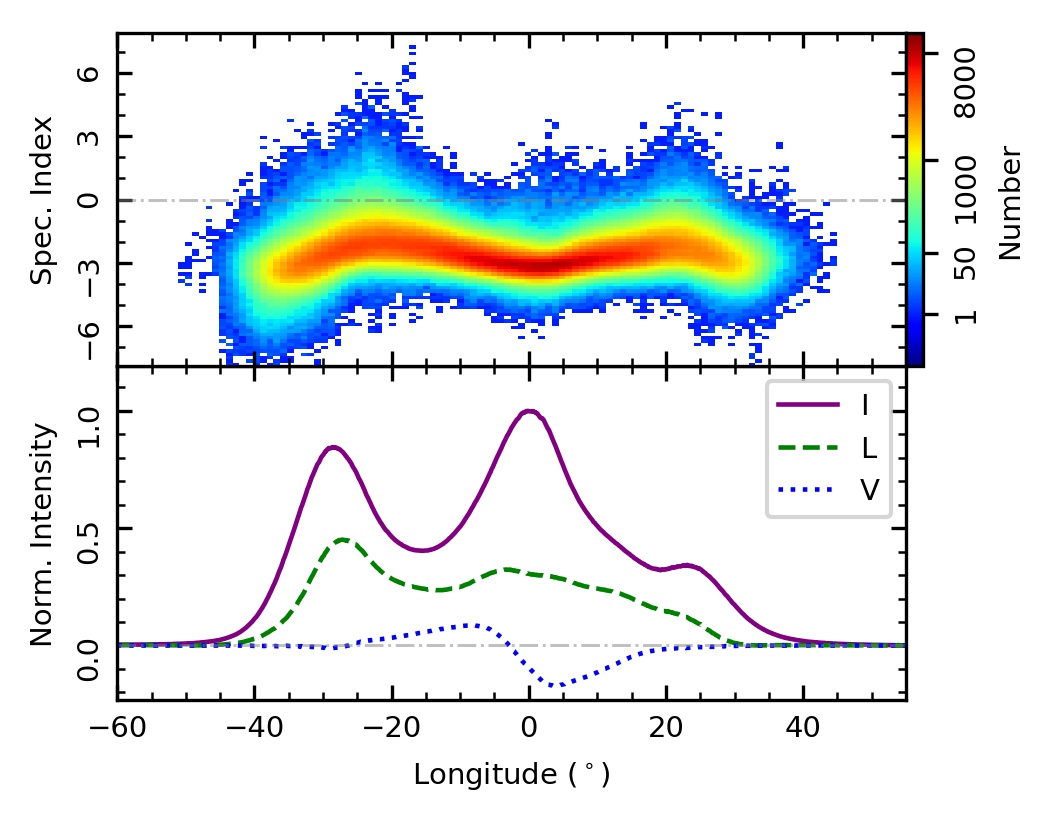}
    \caption{Extended Data Fig.9 $|$ The number distribution of phase-resolved spectral indexes. Data of spectral indexes have an uncertainty less than 0.5 for all individual pulses observed by FAST on 2022-03-08, as shown in the upper subpanel, together with the mean polarization profile for understanding in the lower subpanel scaled with the peak value.}
\end{figure*}

\begin{figure*}
    \centering
    \includegraphics[width=0.8\columnwidth]{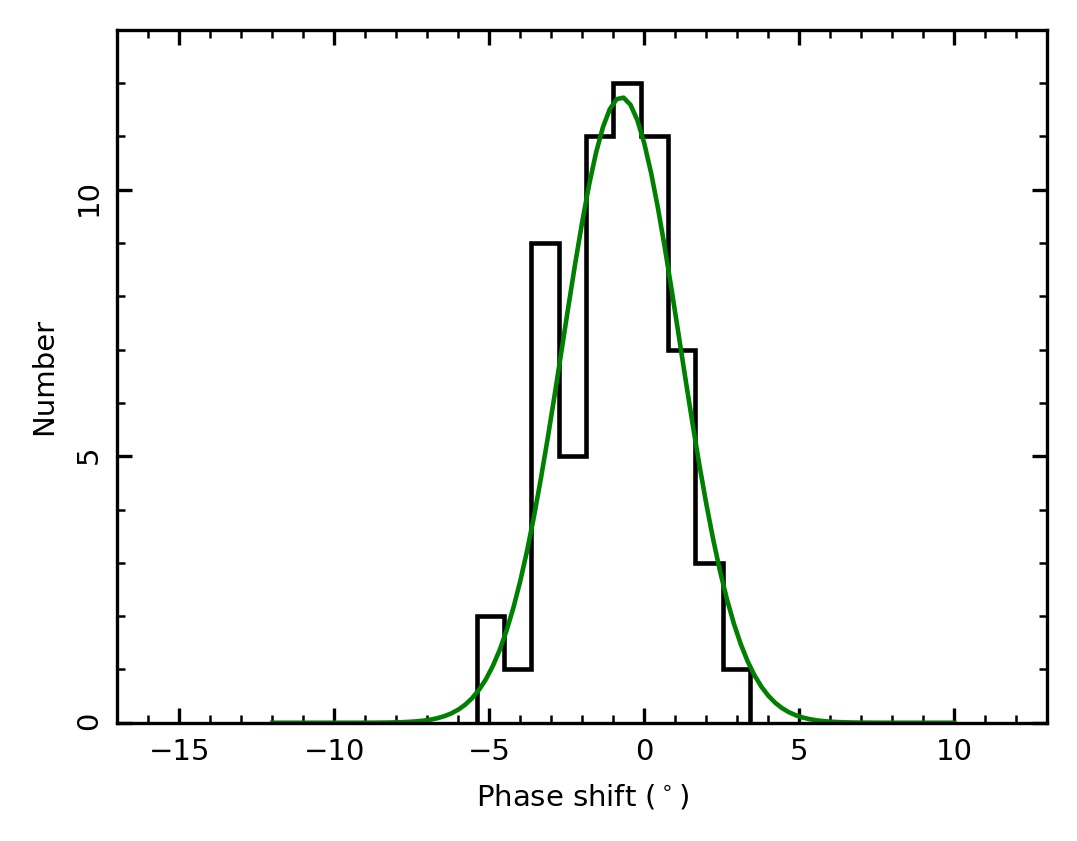}
    \caption{Extended Data Fig.10 $|$ The phase shift distribution of polarization angles of 62 dwarf pulses. The shift values are obtained by comparison of their PA to the mean PA curve at the longitude of dwarf pulses.}
\end{figure*}

\end{document}